\documentclass [aps,prb,showpacs,amsfonts]{revtex4}

\usepackage{epsfig}
\usepackage{bm}

\newcommand{\smfrac}[2]{\mbox{$\frac{#1}{#2}$}}

\begin{document}

\title{Thermoballistic spin-polarized electron transport in paramagnetic
semi\-con\-duc\-tors}

\author{R.\ Lipperheide}
\author{U.\ Wille}
\email{wille@helmholtz-berlin.de}

\affiliation{Helmholtz-Zentrum Berlin f\"{u}r Materialien und Energie
(formerly Hahn-Meitner-Institut Berlin), Lise-Meitner-Campus Wannsee,\\
Glienicker Stra\ss{}e 100, D-14109 Berlin, Germany}

\date{\today}

\begin{abstract}

Spin-polarized electron transport in diluted magnetic semiconductors (DMS) in
the paramagnetic phase is described within the thermoballistic transport model.
In this (semiclassical) model, the ballistic and diffusive transport mechanisms
are unified in terms of a thermoballistic current in which electrons move
ballistically across intervals enclosed between arbitrarily distributed points
of local thermal equilibrium.  The contribution of each interval to the current
is governed by the momentum relaxation length.  Spin relaxation is assumed to
take place during the ballistic electron motion.  In paramagnetic DMS exposed
to an external magnetic field, the conduction band is spin-split due to the
giant Zeeman effect.  In order to deal with this situation, we extend our
previous formulation of thermoballistic spin-polarized transport so as to take
into account an arbitrary (position-dependent) spin splitting of the conduction
band.  The current and density spin pol\-ari\-zations as well as the
magnetoresistance are each obtained as the sum of an equilibrium term
determined by the spin-relaxed chemical potential, and an off-equilibrium
contribution expressed in terms of a spin transport function that is related to
the splitting of the spin-resolved chemical potentials.  The procedures for the
calculation of the spin-relaxed chemical potential and of the spin transport
function are outlined.  As an illustrative example, we apply the
thermoballistic description to spin-polarized transport in DMS/NMS/DMS
heterostructures formed of a nonmagnetic semiconducting sample (NMS) sandwiched
between two DMS layers.  We evaluate the current spin polarization and the
magnetoresistance for this case and, in the limit of small momentum relaxation
length, find our results to agree with those of the standard drift-diffusion
approach to electron transport.\\

\end{abstract}


\maketitle

\section{Introduction}

The electrical injection of spin-polarized carriers from magnetic contacts
into, and their subsequent transport across, nonmagnetic semiconductors (NMS)
constitute outstanding issues in the field of semiconductor spintronics.
\cite{aws02,sch02,zut04,sch05,dya08}  While spin-based semiconductor devices
still await realization, major progress has been achieved over the past years
in the understanding of the conditions and mechanisms governing
polarized-carrier injection and transport.

The low injection efficiency observed for ferromagnetic metal (FM) contacts
found an explanation \cite{sch00} in the large conductivity mismatch between
the contacts and the semiconducting sample.  To circumvent this obstacle, the
introduction of spin-selective interface resistances was suggested
\cite{fil00,ras00} and analyzed theoretically in some detail.
\cite{smi01,fer01,ras02,yuf02,yuf02a,alb02,alb03,kra03}  A considerable
enhancement of the injection efficiency due to externally applied electric
fields was predicted. \cite{yuf02,yuf02a}  On the other hand, the mismatch
problem is mitigated from the outset if contact layers made up of magnetic
semiconductors are used. \cite{zut04,sch05,roy06}  Promisingly high values of
the spin injection efficiency were obtained in experiments
\cite{oes99,fie99,jon00} using paramagnetic (II,Mn)VI DMS contacts.  In the
latter, the conduction band is spin-split due to the giant Zeeman effect
\cite{fur88,die94,cib08} in an external magnetic field, thereby making possible
contact spin polarizations close to 100\%.  For this kind of spin injector, a
novel magnetoresistance effect was observed \cite{sch01,sch04} and
theoretically analyzed \cite{sch01,yuf02a,kha05} within the standard
drift-diffusion theory of electron transport.

Spin-polarized transport in semiconducting structures formed of NMS layers and
paramagnetic (II,Mn)VI DMS layers means electron transmission across
spin-dependent potential profiles.  These profiles depend on position via the
position dependence of the electrostatic potential (composed of the conduction
band edge potential and the external potential) and of the
magnetic-field-induced spin splitting of the conduction band, which changes
abruptly at the NMS/DMS interfaces.  Transport across spin-dependent potential
profiles corresponding to specific combinations of NMS and DMS layers has been
studied theoretically in a number of cases \cite{egu98,guo00,egu01,cha01} in
the (quantum-coherent) ballistic limit.  Recently, the transport properties of
II-VI resonant tunneling devices coupled to paramagnetic DMS contacts were
found \cite{san07,slo07} to depend strongly on the magnitude of the applied
magnetic field.

In the present work, we consider arbitrarily shaped potential profiles
including internal and external electrostatic potentials and exhibiting
arbitrary, position-dependent spin splitting, thereby covering any combination
of NMS layers and paramagnetic DMS layers that may occur in semiconducting
structures.  We study spin-polarized electron transport within the
(semiclassical) thermoballistic description \cite{lip03} of carrier transport
in nondegenerate semiconductors.  The basic element of this description is the
thermoballistic current, in which electrons move ballistically across intervals
enclosed between points of local thermal equilibrium.  The contribution to the
current of each such "ballistic interval" is governed by the momentum
relaxation length ("mean free path").  The thermoballistic transport mechanism
is intermediate between the diffusive (Drude) process, where the electrons move
from one state of equilibrium to another, infinitesimally close-lying state
(mean free path tending toward zero), and, on the other hand, the ballistic,
collision-free motion across the electrostatic potential profile (mean free
path tending toward infinity).  In the thermoballistic current, diffusive and
ballistic transport are thus unified in a way that allows the effect of these
two aspects of the transport mechanism to be studied.  In Ref.\
\onlinecite{lip05}, we have introduced spin relaxation into the thermoballistic
description for the case of electron transport in NMS.  There, we have
disregarded any spin splitting of the conduction band.  Applications
\cite{lip05,lip06} have dealt with spin-polarized transport in heterostructures
formed of an NMS layer sandwiched between two ferromagnetic metal contacts.  In
the present paper, we put forward the systematic extension of the
thermoballistic approach to spin-polarized electron transport across {\em
spin-split} potential profiles.  While we are mainly concerned here with the
general formulation of this extension, we also treat, within a simplified
picture, spin-polarized transport in heterostructures formed of an NMS layer
enclosed between two DMS layers.  Applications to specific experimental
configurations will be deferred to future work.

As a prerequisite to the thermoballistic description, we formulate, in the next
two sections, the details of {\em ballistic} spin-polarized electron transport
across spin-split potential profiles.  In Sec.~II, we introduce the
electron densities at the points of local thermal equilibrium enclosing
a ballistic interval, and construct the currents injected from these points
into that interval.  In Sec.~III, we continue the injected currents and the
densities from the points of local thermal equilibrium into the ballistic
interval, and introduce the balance equation that describes spin relaxation
inside the interval.  The off-equilibrium ballistic spin-polarized current and
density are expressed in terms of a spin transport function that is related to
the splitting of the spin-resolved chemical potentials.  In Sec.~IV, the {\em
thermoballistic} currents and densities are constructed by summing up the
contributions from all ballistic intervals.  The magnitude of these
contributions is governed by the momentum relaxation length.  The current and
density spin polarizations as well as the magnetoresistance are each expressed
as the sum of an equilibrium term determined by the spin-relaxed chemical
potential, and an off-equilibrium contribution given in terms of the spin
transport function.  The spin-relaxed chemical potential is described through a
resistance function.  The latter, as well as the spin transport function, are
each calculated from an integral equation.  Section V deals with the
thermoballistic description of spin-polarized electron transport in DMS/NMS/DMS
heterostructures.  We derive explicit expressions for the current spin
polarization and the magnetoresistance and compare these, in the limit of small
momentum relaxation length, to those of the standard drift-diffusion approach
to electron transport.  Numerical results for DMS/NMS/DMS heterostructures are
presented and discussed in Sec.~VI.  In Sec.~VII, we summarize the contents of
this paper and make some concluding remarks.

\begin{figure}[t]
\includegraphics[width=0.7\textwidth]{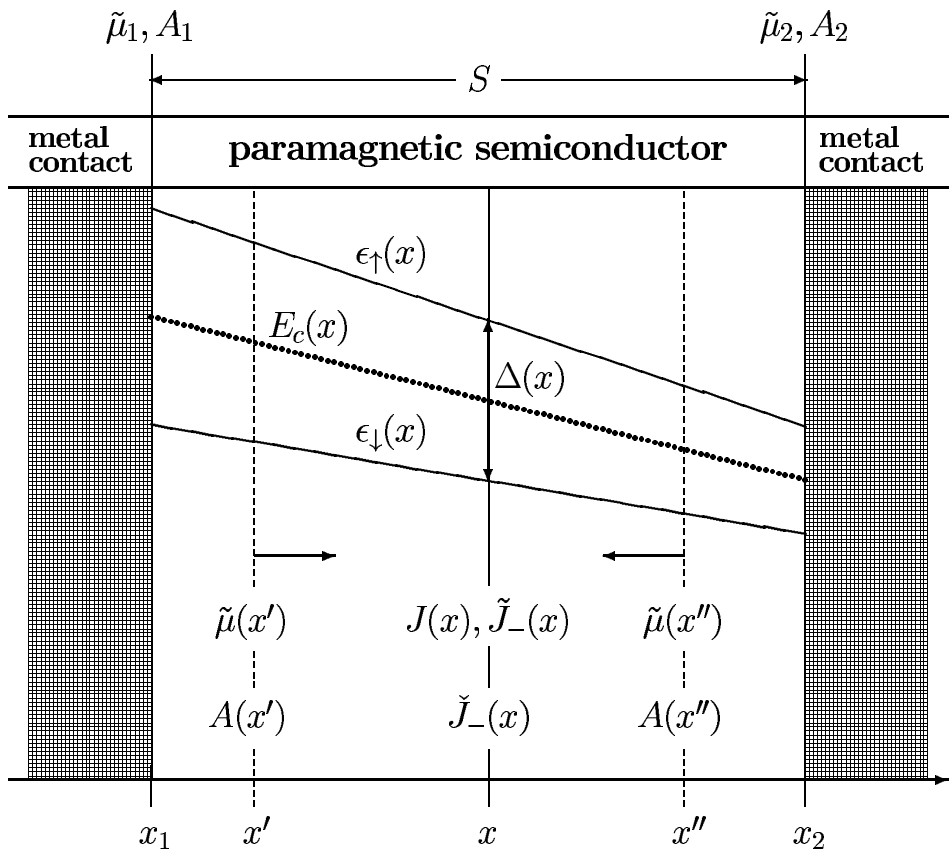}
\caption{Schematic diagram showing, in a one-dimensional geometry, a
paramagnetic semiconducting sample of length $S$ enclosed between two
nonmagnetic metal contacts.  The spin splitting $\Delta (x)$ of the potential
profile $E_{c}(x)$ gives rise to the spin-dependent potential profiles
$\epsilon_{\uparrow , \downarrow}(x)$.  The coordinates $x'$ and $x''$ denote
points of local thermal equilibrium.  Electrons injected at $x' (x'')$ toward
the right (left) move ballistically across the profiles $\epsilon_{\uparrow ,
\downarrow}(x)$ to reach the end-point $x'' (x')$ of the ballistic transport
interval $[x',x'']$.  The thermoballistic equilibrium electron currents
$J(x)$ and $\tilde{J}_{-}(x)$ are constructed from the common spin-relaxed
chemical potential $\tilde{\mu} (x')$ [with boundary values
$\tilde{\mu}_{1,2}$ at the contact/semiconductor interfaces] according to
Eq.~(\ref{eq:64}).  Correspondingly, the off-equilibrium thermoballistic
current $\check{J}_{-}(x)$ is constructed from the spin transport function
$A(x')$ [with boundary values $A_{1,2}$].
}
\label{fig:1}
\end{figure}

\section{Points of thermal equilibrium:\ Densities and injected currents}

The thermoballistic description of spin-polarized electron transport in NMS
\cite{lip05} makes use, in a one-dimensional geometry, of ballistic electron
currents and densities in ``ballistic intervals'' between two points of local
thermal equilibrium, $x'$ and $x''$ ($x_1 \leq x' < x'' \leq x_2$); see
Fig.~\ref{fig:1}.  The ballistic electron transport across such an interval is
determined by the densities at the points of local thermal equilibrium and by
the shape of the potential profile inside the interval.  For the purpose of
generalizing the thermoballistic approach to spin-polarized electron transport
in paramagnetic DMS, we introduce the thermal-equilibrium densities
$n_{\uparrow , \downarrow}(x')$ and $n_{\uparrow , \downarrow}(x'')$ for
spin-up $(\uparrow)$ and spin-down $(\downarrow)$ conduction band states, as
well as the spin-dependent potential profiles
\begin{equation}
\epsilon_{\uparrow , \downarrow}(x) = E_{c}(x) \pm \Delta (x) /2
\label{eq:01}
\end{equation}
at position $x \in [x_{1},x_{2}]$.  Here, the (spin-independent) potential
$E_c(x)$ comprises the conduction band edge potential and the external
electrostatic potential, and $\Delta(x)$ is the Zeeman splitting of the
conduction band due to an external magnetic field \cite{fur88} [we restrict
ourselves to considering a single Landau level whose energy is assumed to
be included in $E_{c}(x)$].  In developing our formalism, we assume both
$E_{c}(x)$ and $\Delta(x)$, and hence $\epsilon_{\uparrow , \downarrow}(x)$, to
be continuous functions of $x$ in the interval $[x_{1}, x_{2}]$.  Abrupt
changes in one or the other of these functions, which occur at the interfaces
in heterostructures, may be described, in a simplified picture, in terms of
discontinuous functions. This aspect will be illustrated in Sec.~V below for
the case of DMS/NMS/DMS heterostructures.

In the thermoballistic transport mechanism, momentum relaxation takes place
exclusively and instantaneously at the points of local thermal equilibrium.
Spin relaxation via spin-flip scattering processes, on the other hand, occurs
during the electron motion across the ballistic intervals (since the electrons
spend only an infinitesimally short time span at the points of local thermal
equilibrium, it is only inside the ballistic intervals that they can experience
spin relaxation).  This separation of momentum and spin relaxation is in
accordance with the D'yakonov-Perel' relaxation mechanism \cite{dya71,dya08a},
which is known \cite{zut04,dya08} to be the dominant relaxation mechanism at
large donor doping levels and at high temperatures.  Therefore, our approach
covers a broad range of cases interesting from the point of view of experiment.

\subsection{Electron densities}

The electron motion in a ballistic interval is activated at a point of local
thermal equilibrium at one end and terminated at another such point at the
other end.  In the semiclassical model, the points of local thermal equilibrium
are characterized by a local chemical potential ("quasi-Fermi level").  In a
nondegenerate system, the spin-resolved electron densities,
$n_{\uparrow , \downarrow}(x')$, are given by
\begin{equation}
n_{\uparrow , \downarrow}(x') = \frac{N_{c}}{2} \,
e^{- \beta [\epsilon_{\uparrow , \downarrow}(x') - \mu_{\uparrow ,
\downarrow}(x')]} \, ,
\label{eq:02}
\end{equation}
where $\mu_{\uparrow , \downarrow}(x')$ are the spin-resolved chemical
potentials, $N_c/2 = (2\pi m^*/\beta h^2)^{3/2}$ is the effective density of
states of either spin at the conduction band edge, $m^*$ is the effective mass
of the electrons (which, for simplicity, is assumed to be independent of
position and of the external magnetic field), and $\beta = (k_B T)^{-1}$.  In
the spin-relaxed state, the potentials $\mu_{\uparrow , \downarrow}(x')$ become
equal to the common "spin-relaxed chemical potential" $\tilde{\mu}(x')$.  Thus,
the spin-relaxed densities at the point of thermal equilibrium $x'$, denoted by
$\tilde{n}_{\uparrow , \downarrow}(x')$, are
\begin{equation}
\tilde{n}_{\uparrow , \downarrow}(x') = \frac{N_{c}}{2} \, e^{- \beta [
\epsilon_{\uparrow , \downarrow}(x') - \tilde{\mu}(x')]} \, .
\label{eq:03}
\end{equation}
For the Boltzmann factors $e^{-\beta \epsilon_{\uparrow , \downarrow}(x')}$, we
introduce the notation
\begin{equation}
\vartheta_{\uparrow ,  \downarrow}(x') =
e^{-\beta \epsilon_{\uparrow , \downarrow}(x')}
\label{eq:03a}
\end{equation}
and define
\begin{equation}
\vartheta_{\pm}(x') = \vartheta_{\uparrow}(x') \pm \vartheta_{\downarrow}(x')
\label{eq:03b}
\end{equation}
(the $``\pm''$ notation will be used {\em mutatis mutandis} for various other
quantities appearing further below), so that we have
\begin{equation}
\tilde{n}_{\pm}(x') = \frac{N_{c}}{2} \, \vartheta_{\pm}(x') \, e^{ \beta
\tilde{\mu}(x')} \, .
\label{eq:03bc}
\end{equation}
The total (i.e.,\ spin-summed) spin-relaxed density, $\tilde{n}(x') \equiv
\tilde{n}_{+}(x')$, is
\begin{equation}
\tilde{n}(x') = \frac{N_{c}}{2} \, \vartheta_{+}(x') \, e^{ \beta
\tilde{\mu}(x')} \, ,
\label{eq:03c}
\end{equation}
and for the ``static'' spin polarization $P(x')$ of the conduction band
electrons, i.e., the polarization in the spin-relaxed state at zero external
{\em electric} field, we have
\begin{equation}
P(x') \equiv \frac{\tilde{n}_-(x')}{ \tilde{n}(x')} =
\frac{\vartheta_-(x')}{\vartheta_+(x')} = - \tanh \bm{(} \beta \Delta(x')/2
\bm{)} \, .
\label{eq:06}
\end{equation}
We note the relation
\begin{equation}
\vartheta_{\uparrow , \downarrow}(x') = \smfrac{1}{2} \, [1 \pm
P(x')] \, \vartheta_+(x')
\label{eq:07}
\end{equation}
expressing the Boltzmann factors $\vartheta_{\uparrow , \downarrow}(x')$ in
terms of $P(x')$.

For the dynamical description of spin-polarized transport, we now introduce the
ratios $\alpha_{\uparrow , \downarrow}(x')$ of the densities (\ref{eq:02}) to
their spin-relaxed counterparts (\ref{eq:03}),
\begin{equation}
\alpha_{\uparrow , \downarrow}(x') \equiv  \frac{n_{\uparrow ,
\downarrow}(x')}{\tilde{n}_{\uparrow , \downarrow}(x')} = e^{\beta
[\mu_{\uparrow , \downarrow}(x') -  \tilde{\mu}(x')]} \, ;
\label{eq:08}
\end{equation}
working with these quantities will provide us with a description in terms of
linear equations, instead of the nonlinear description that we would encounter
when employing the chemical potentials $\mu_{\uparrow , \downarrow}(x')$
themselves.  This aspect has been emphasized previously \cite{yuf02a} within
the diffusive approach to spin polarization in nondegenerate semiconductors.
[We note that the quantity $\alpha_{\uparrow , \downarrow}(x')$ introduced in
Eq.~(\ref{eq:08}) is different from that of Eq.~(3.8) in
Ref.~\onlinecite{lip05}; in the spin-relaxed state, we have in the present case
$\alpha_{\uparrow , \downarrow}(x') = 1$, while in Ref.~\onlinecite{lip05} we
have $\alpha_{\uparrow , \downarrow}(x') = 1/2$.] Using Eqs.~(\ref{eq:03a}),
(\ref{eq:03c}), (\ref{eq:07}), and (\ref{eq:08}), we can write the densities
$n_{\uparrow , \downarrow}(x')$ in the form
\begin{equation}
n_{\uparrow , \downarrow}(x') = \smfrac{1}{2} \, \tilde{n}(x') \,
[1 \pm P(x')] \, \alpha_{\uparrow , \downarrow}(x') \, ,
\label{eq:09}
\end{equation}
so that
\begin{equation}
n_{\pm}(x') = \smfrac{1}{2} \, \tilde{n}(x') \,
[\alpha_{\pm}(x') + P(x')\,\alpha_{\mp}(x')] \, .
\label{eq:17}
\end{equation}
Here, the functions $\alpha_{+}(x')$ and $\alpha_{-}(x')$ appear as independent
dynamical quantities.

We now require the total density, $n(x') \equiv n_{+}(x')$, to be unaffected by
spin relaxation, i.e., we set
\begin{equation}
n(x') = \tilde{n}(x') \, .
\label{eq:10}
\end{equation}
This requirement is justified \cite{yuf02} for n-doped (unipolar) systems, for
which spin-flip scattering processes involving valence band states can be
disregarded.  Substituting Eq.~(\ref{eq:10}) in the upper Eq.~(\ref{eq:17}),
we find $\alpha_{+}(x')$ and $\alpha_{-}(x')$ related by
\begin{equation}
\alpha_{+}(x') = 2 - P(x') \alpha_{-}(x') \,  ,
\label{eq:17a}
\end{equation}
so that we can eliminate $\alpha_{+}(x')$ from the lower Eq.~(\ref{eq:17})
to obtain the spin-polarized density, $n_{-}(x')$, as
\begin{equation}
n_{-}(x') = n(x') \, \{ P(x') + \smfrac{1}{2}[Q(x')]^2 \alpha_{-}(x')\} \, ,
\label{eq:17b}
\end{equation}
where
\begin{equation}
[Q(x')]^{2} \equiv 1 - [P(x')]^{2} = \frac{4
\vartheta_{\uparrow}(x') \vartheta_{\downarrow}(x')}{[\vartheta_{+}(x')]^2 }
= \frac{1}{\cosh^{2} \bm{(} \beta \Delta(x')/2 \bm{)} } \, .
\label{eq:17c}
\end{equation}
In the spin-relaxed state, we have
\begin{equation}
\alpha_{\uparrow , \downarrow}(x')  = 1 \, ,
\label{eq:15a}
\end{equation}
and hence
\begin{equation}
\alpha_{-}(x') = 0 \, , \, \,  \alpha_{+}(x') = 2 \, .
\label{eq:15b}
\end{equation}
Then, from Eq.~(\ref{eq:17b}), the equilibrium spin-polarized density at the
point of thermal equilibrium $x'$, $\tilde{n}_{-}(x')$, is
\begin{equation}
\tilde{n}_{-}(x') =  n(x') \, P(x')
\label{eq:15c}
\end{equation}
[see Eq.~(\ref{eq:06})], and the off-equilibrium spin-polarized density,
$\check{n}_{-}(x')$, is there
\begin{equation}
\check{n}_{-}(x') \equiv n_{-}(x') - \tilde{n}_{-}(x') = \smfrac{1}{2}
n(x') \, [Q(x')]^2 \alpha_{-}(x') \, .
\label{eq:15d}
\end{equation}
The off-equilibrium spin-polarized density is proportional to the quantity
$\alpha_-(x')$, i.e., to the difference of the spin-resolved densities relative
to their spin-relaxed values [see Eq.~(\ref{eq:08})], but also contains, via
the factor $[Q(x')]^2$, the effect of the static polarization $P(x')$.

For later use, we establish the relation between the difference of the
spin-resolved chemical potentials, $\mu_{-}(x')$, and the quantity
$\alpha_{-}(x')$ which will appear as the key element determining the
spin dynamics of off-equilibrium spin-polarized transport in the
thermoballistic approach (see Secs.~III.B and IV.C).  From Eq.~(\ref{eq:10}),
we obtain, using Eqs.~(\ref{eq:02}), (\ref{eq:03c}), and (\ref{eq:08}),
\begin{equation}
\vartheta_{\uparrow}(x') \alpha_{\uparrow}(x') +
\vartheta_{\downarrow}(x')\alpha_{\downarrow}(x') = \vartheta_+(x') \, ,
\label{eq:10a}
\end{equation}
and hence, using Eq.~(\ref{eq:07}),
\begin{equation}
\alpha_{\uparrow , \downarrow}(x') = 1 \pm \smfrac{1}{2}[1 \mp P(x')]
\alpha_{-}(x') \, .
\label{eq:10b}
\end{equation}
With the help of Eq.~(\ref{eq:08}), we then have
\begin{equation}
\mu_{-}(x') = \frac{1}{\beta} \, \ln \left(
\frac{\alpha_{\uparrow}(x')}{\alpha_{\downarrow}(x')} \right) =
\frac{1}{\beta} \, \ln \left( \frac{1 +  \frac{1}{2}[1 -
P(x')] \, \alpha_{-}(x')}{1 - \frac{1}{2} [1 + P(x')] \, \alpha_{-}(x')}
\right) ,
\label{eq:16}
\end{equation}
and, reversely,
\begin{equation}
\alpha_{-}(x') = \frac{2 \tanh \bm{(} \beta \mu_-(x')/2 \bm{)} }{1+ P(x') \tanh
\bm{(} \beta  \mu_-(x')/2 \bm{)} } \, .
\label{eq:16a}
\end{equation}
For the {\em mean chemical potential} $\bar{\mu}(x')\equiv \frac{1}{2}
\mu_+(x')$, we find
\begin{eqnarray}
\bar{\mu}(x') &=& \tilde{\mu}(x') + \frac{1}{2\beta} \, \ln \bm{(}
\alpha_{\uparrow}(x')\alpha_{\downarrow}(x') \bm{)} \nonumber \\ &=&
\tilde{\mu}(x') + \frac{1}{2\beta} \, \ln \bm{(} \{ 1 +
\smfrac{1}{2}[1 - P(x')] \, \alpha_{-}(x') \} \{ 1 - \smfrac{1}{2}
[1 + P(x')] \, \alpha_{-}(x') \} \bm{)} \nonumber \\ &=&
\tilde{\mu}(x') + \frac{1}{2\beta} \, \ln \bm{(}  1 - P(x') \alpha_{-}(x') -
\smfrac{1}{4}[Q(x')]^{2} [\alpha_{-}(x')]^{2} \bm{)} \, .
\label{eq:16aa}
\end{eqnarray}
Finally, we quote the relation between the spin-resolved chemical potentials
$\mu_{\uparrow , \downarrow}(x')$ and the common spin-relaxed chemical
potential $\tilde{\mu}(x')$,
\begin{equation}
\smfrac{1}{2}[1+P(x')]e^{\beta\mu_{\uparrow}(x')} + \smfrac{1}{2}
[1-P(x')] e^{\beta\mu_{\downarrow}(x')} = e^{\beta \tilde{\mu}(x')}
\, ,
\label{eq:16b}
\end{equation}
which, like Eq.~(\ref{eq:10a}), follows from Eq.~(\ref{eq:10}), but this time
using Eqs.~(\ref{eq:02}), (\ref{eq:03a}), (\ref{eq:03c}), and (\ref{eq:07}).

\subsection{Current injection}

In the thermoballistic description, a point of thermal equilibrium lies between
two ballistic intervals (except at the ends of the sample, which will be
considered further below), and electrons from the left- and right-lying
intervals enter into it to be equilibrated instantaneously.  At the same time,
equilibrated electrons, which have no preferred direction of motion, are
injected symmetrically (i.e.,\ "half-and-half") into either ballistic interval,
forming the spin-resolved electron current densities (electron currents, for
short) $v_e n_{\uparrow , \downarrow}(x')$, where $v_e = (2\pi
m^*\beta)^{-1/2}$ is the emission velocity.  If, however, there lies a
potential barrier $\epsilon^m_{\uparrow , \downarrow}(x',x'') -
\epsilon_{\uparrow , \downarrow}(x') > 0$ $( \epsilon^m_{\uparrow ,
\downarrow}(x',x'')$ is the maximum of the spin-dependent potential profile in
the interval $[x',x''] )$ between, say, the left point of current injection,
$x'$, and the opposite point of current absorption, $x''$, part of the current
will be reflected at this barrier (and equilibrated while it travels back and
forth between $x'$ and the left side of the barrier, -- we may call it the
``confined current'').  The part that surmounts the barrier (henceforth called
the ``injected current'') is given by
\begin{eqnarray}
J^{l}_{\uparrow , \downarrow}(x',x'') &=& \frac{N_{c}}{2} \, (2 \beta
/\pi m^{\ast})^{1/2} \int_{0}^{\infty} dp \,
\frac{p}{m^{\ast}} \, e^{-\beta [p^{2}/2m^{\ast} +
\epsilon_{\uparrow , \downarrow}(x') - \mu_{\uparrow , \downarrow}(x')]}
\, \Theta (p^{2}/2m^{\ast} + \epsilon_{\uparrow , \downarrow}(x') -
\epsilon_{\uparrow , \downarrow}^{m}(x',x'')) \nonumber \\ &=& v_e
n_{\uparrow , \downarrow}(x') \, T^l_{\uparrow , \downarrow}(x',x'') \, ,
\label{eq:17d}
\end{eqnarray}
where
\begin{equation}
T^l_{\uparrow , \downarrow}(x',x'') \equiv e^{- \beta [ \epsilon^{m}_{\uparrow
, \downarrow}(x',x'') - \epsilon_{\uparrow , \downarrow}(x')]} =
\frac{\vartheta^m_{\uparrow , \downarrow}(x',x'')}{\vartheta_{\uparrow ,
\downarrow}(x')}
\label{eq:18}
\end{equation}
is the (classical) transmission probability for the spin-resolved current
injected at the left end-point $x'$ of the ballistic interval to reach the
opposite right end-point $x''$.  Here, we have introduced the notation
\begin{equation}
\vartheta^m_{\uparrow , \downarrow}(x',x'') =
e^{-\beta \epsilon^m_{\uparrow , \downarrow}(x',x'')} \, ,
\label{eq:19a}
\end{equation}
which corresponds to the definition (\ref{eq:03a}) for the Boltzmann factors
$\vartheta_{\uparrow , \downarrow}(x')$, with $\epsilon_{\uparrow ,
\downarrow}(x')$ replaced with $\epsilon^m_{\uparrow , \downarrow}(x',x'')$
[note that the factors $\vartheta^m_{\uparrow}(x',x'')$ and
$\vartheta^m_{\downarrow}(x',x'')$ are {\em local} functions depending, for a
given ballistic interval $[x',x'']$, each on the position of the maximum of
the potential profiles $\epsilon_{\uparrow}(x)$ and $\epsilon_{\downarrow}(x)$,
respectively].  If the potential profile is constant along the interval or if
its maximum lies at the injection point $x'$ itself, then
$\epsilon^{m}_{\uparrow , \downarrow}(x',x'') = \epsilon_{\uparrow ,
\downarrow}(x')$ and $T^l_{\uparrow , \downarrow}(x',x'') = 1$.

In analogy to Eq.~(\ref{eq:09}) for the spin-resolved densities, we now write
the currents (\ref{eq:17d}) in the form
\begin{equation}
J^{l}_{\uparrow , \downarrow}(x',x'') = \smfrac{1}{2} \tilde{J}^{l}(x',x'') \,
[1 \pm P^m(x',x'')] \, \alpha_{\uparrow , \downarrow}(x') \, ,
\label{eq:19}
\end{equation}
where
\begin{equation}
\tilde{J}^{l}(x',x'') \equiv \tilde{J}^{l}_{+}(x',x'') = \frac{v_e N_c}{2}
\vartheta^m_+(x',x'') e^{\beta \tilde{\mu}(x')}
\label{eq:25a}
\end{equation}
is the total spin-relaxed current injected at $x'$ into the interval $[x',x'']$
from the left, and
\begin{equation}
P^{m}(x',x'') = \frac{\vartheta_{-}^{m}(x',x'')}{\vartheta_{+}^{m}(x',x'')}
\label{eq:25b}
\end{equation}
is a static, ``nonlocal" spin polarization depending on the (generally
different) positions of the maximum of the potential profile for the spin-up
state and for the spin-down state, respectively [compare with the analogous
relation (\ref{eq:06}) for the {\em local} spin polarization $P(x')$].

From expression (\ref{eq:19}), we obtain, in analogy to Eq.~(\ref{eq:17}),
\begin{equation}
J^{l}_{\pm}(x',x'') = \smfrac{1}{2} \, \tilde{J}^{l}(x',x'')
\, [\alpha_{\pm}(x') + P^{m}(x',x'') \, \alpha_{\mp}(x')]
\, .
\label{eq:27}
\end{equation}
In line with the condition (\ref{eq:10}) imposed on the total density $n(x')$,
we now require the total current $J^{l}(x',x'') \equiv J^{l}_{+}(x',x'')$ to
equal its spin-relaxed limit,
\begin{equation}
J^{l}(x',x'') =  \tilde{J}^{l}(x',x'') \, .
\label{eq:27b}
\end{equation}
Following the argument leading to expression (\ref{eq:17b}) for the
spin-polarized density $n_{-}(x')$, we then find
\begin{equation}
J^{l}_{-}(x',x'') =  J^{l}(x',x'') \, \{ P^m(x',x'') +
\smfrac{1}{2}[Q^m(x',x'')]^2 \alpha_{-}(x') \} \, ,
\label{eq:27a}
\end{equation}
where
\begin{equation}
[Q^m(x',x'')]^2 \equiv 1-[P^m(x',x'')]^2 = \frac{4
\vartheta_{\uparrow}^{m}(x',x'') \vartheta_{\downarrow}^{m}(x',x'')
}{[\vartheta_{+}^{m}(x',x'')]^{2}}
\label{eq:27aa}
\end{equation}
(see the analogous relation (\ref{eq:17c}) for $[Q(x')]^2$).

In the spin-relaxed state [see Eq.~(\ref{eq:15b})], Eq.~(\ref{eq:27}) yields
for the injected equilibrium spin-polarized current
\begin{equation}
\tilde{J}^{l}_{-}(x',x'') = J^{l}(x',x'')\,  P^m(x',x'') \, ,
\label{eq:27bb}
\end{equation}
so that the injected off-equilibrium spin-polarized current is
\begin{equation}
\check{J}^l_{-}(x',x'') \equiv J^l_{-}(x',x'') - \tilde{J}^l_{-}(x',x'')
= \smfrac{1}{2} J^l(x',x'')  \, [Q^m(x',x'')]^2 \alpha_{-}(x') \, ,
\label{eq:27c}
\end{equation}
in parallel to Eqs.~(\ref{eq:15c}) and (\ref{eq:15d}), respectively.

All the preceding formulas hold also in the case where the left end-point of
the ballistic interval coincides with the left end-point of the sample, $x' =
x_1$, and similarly for the right end-points, $x'' = x_2$, with the
understanding that the chemical potentials $\tilde{\mu}(x_1) = \tilde{\mu}_1$
and $\tilde{\mu}(x_2) = \tilde{\mu}_2$ have fixed values given by the chemical
potentials in the contacts bordering on the semiconducting sample at either
end.

\section{Spin-polarized transport across a ballistic interval}

The electron currents discussed up to now are the currents injected into the
ballistic interval at the point of thermal equilibrium {\em as they leave the
latter}.  Once they have entered this interval, they become {\em transmitted
ballistic currents}, which will be considered in the following.

\subsection{Continuing the injected currents and densities into the
ballistic interval}

We again consider injection at the left point of thermal equilibrium, $x'$, of
the ballistic interval $[x',x'']$.  If the system is in the spin-relaxed state
at the injection point $x'$, we have $\mu_{\uparrow , \downarrow}(x') =
\tilde{\mu}(x')$ and $\alpha_{\uparrow , \downarrow}(x') = 1$.  Then the
transmitted (in the following, we omit the attribute ``transmitted'') ballistic
spin-resolved current at $x'$, given by expression (\ref{eq:19}), propagates
along the ballistic interval without spin relaxation, i.e., it is conserved.
If, however, $\mu_{\uparrow , \downarrow}(x') \neq \tilde{\mu}(x')$, the
injected current will relax along the ballistic interval and hence will no
longer be conserved.  It is then natural to continue expression (\ref{eq:19})
into the ballistic interval $[x',x'']$ by replacing $\alpha_{\uparrow ,
\downarrow}(x')$ with a more general quantity $\alpha^l_{\uparrow ,
\downarrow}(x',x'';x)$ depending both on the end-points $x'$ and $x''$ {\em
and} on the position $x \in [x',x'']$, so that the ballistic spin-resolved
current is obtained as
\begin{equation}
J^{l}_{\uparrow , \downarrow}(x',x'';x) = \smfrac{1}{2} J^{l}(x',x'') \, [1 \pm
P^m(x',x'')] \, \alpha^{l}_{\uparrow , \downarrow}(x',x'';x) =
\frac{v_{e}N_{c}}{2} \, \vartheta^{m}_{\uparrow , \downarrow}(x',x'') \,
e^{\beta \tilde{\mu}(x')} \, \alpha^{l}_{\uparrow , \downarrow}(x',x'';x) \, ,
\label{eq:29}
\end{equation}
with the initial condition
\begin{equation}
\alpha^{l}_{\uparrow , \downarrow}(x',x'';x') = \alpha_{\uparrow ,
\downarrow}(x') \, ,
\label{eq:30}
\end{equation}
which holds for all $x' \geq x_1$.

In line with the density ratios $\alpha_{\uparrow , \downarrow}(x')$ in
Eq.~(\ref{eq:19}), the generalized quantities $\alpha^{l}_{\uparrow ,
\downarrow}(x',x'';x)$ express the deviation of the currents inside the
ballistic interval from their spin-relaxed values, so that here the condition
for spin equilibrium is again
\begin{equation}
\alpha^{l}_{\uparrow , \downarrow}(x',x'';x) = 1
\label{eq:37aa}
\end{equation}
[see Eq.~(\ref{eq:15a})], and hence
\begin{equation}
\alpha^{l}_{-}(x',x'';x) = 0 \; , \; \alpha^{l}_{+}(x',x'';x) = 2 \, ,
\label{eq:37bb}
\end{equation}
as in Eq.~(\ref{eq:15b}).

Extending condition (\ref{eq:27b}) and the procedure ensuing therefrom, we
obtain for the (conserved) total ballistic current {\em inside} the ballistic
interval
\begin{equation}
J^{l}(x',x'';x) \equiv J_{+}^{l}(x',x'';x) = J^{l}(x',x'') =
\frac{v_e  N_c}{2} \vartheta^m_+(x',x'') e^{\beta \tilde{\mu}(x')}
\label{eq:37b}
\end{equation}
and, similarly, for the (conserved) ballistic equilibrium  spin-polarized
current
\begin{equation}
\tilde{J}^{l}_{-}(x',x'';x) = J^{l}(x',x'')\,  P^m(x',x'') = \frac{v_e
N_c}{2} \vartheta^m_-(x',x'') e^{\beta \tilde{\mu}(x')}  \,
\label{eq:37c}
\end{equation}
[see Eq.~(\ref{eq:27bb})], while for the ballistic off-equilibrium
spin-polarized current we have
\begin{eqnarray}
\check{J}^l_{-}(x',x'';x) &=& \smfrac{1}{2} J^l(x',x'') \,
[Q^m(x',x'')]^2 \alpha_{-}^{l}(x',x'';x) \nonumber \\ &=& \frac{v_{e} N_{c}}{2}
\vartheta^{m}(x',x'') e^{\beta \tilde{\mu}(x')} \alpha_{-}^{l}(x',x'';x)
\label{eq:37d}
\end{eqnarray}
[see Eq.~(\ref{eq:27c})], where
\begin{equation}
\vartheta^{m}(x',x'') = \frac{2 \vartheta_{\uparrow}^{m}(x',x'')
\vartheta_{\downarrow}^{m}(x',x'')}{\vartheta_{+}^{m}(x',x'')} =
\smfrac{1}{2} \vartheta_{+}^{m}(x',x'') [Q^{m}(x',x'')]^{2}  \, .
\label{eq:37e}
\end{equation}
In Eq.~(\ref{eq:37d}), the {\em spin dynamics} of the ballistic off-equilibrium
spin-polarized current $\check{J}^l_{-}(x',x'';x)$ is determined by the factor
$\alpha_{-}^{l}(x',x'';x)$, whose dependence on the coordinate $x$ reflects the
spin relaxation inside the ballistic interval $[x',x'']$.  This dependence will
be considered in Sec.~III.B.

The density $n^{l}_{\uparrow , \downarrow}(x',x'';x)$ associated with the
ballistic spin-resolved current $J^{l}_{\uparrow , \downarrow}(x',x'';x)$ will
be called ``ballistic spin-resolved density''.  It is obtained from
Eq.~(\ref{eq:29}) by replacing $v_{e}$ with
$C_{\uparrow , \downarrow}(x',x'';x)/2$,
\begin{equation}
n^{l}_{\uparrow , \downarrow}(x',x'';x) = \frac{N_c}{4}
D^m_{\uparrow , \downarrow}(x',x'';x) \, e^{\beta \tilde{\mu}(x')}
\,\alpha^{l}_{\uparrow , \downarrow}(x',x'';x)
\label{eq:31}
\end{equation}
(see the analogous relation between the current (2.2) and the density (2.7) in
Ref.~\onlinecite{lip05}). Here,
\begin{equation}
D^m_{\uparrow , \downarrow}(x',x'';x) = C_{\uparrow , \downarrow}(x',x'';x)
\, \vartheta^m_{\uparrow , \downarrow}(x',x'')
\label{eq:42}
\end{equation}
and
\begin{equation}
C_{\uparrow , \downarrow}(x',x'';x) = \frac{1}{T_{\uparrow ,
\downarrow}(x',x'';x)}
\, {\rm erfc}\bm{(}[- \ln T_{\uparrow , \downarrow}(x',x'';x)]^{1/2} \bm{)}
\label{eq:33}
\end{equation}
[see Eq.~(2.8) of  Ref.~\onlinecite{lip05}], with
\begin{equation}
T_{\uparrow , \downarrow}(x',x'';x) \equiv e^{- \beta [ \epsilon^{m}_{\uparrow
, \downarrow}(x',x'') - \epsilon_{\uparrow , \downarrow}(x)]} =
\frac{\vartheta^m_{\uparrow ,\downarrow}(x',x'')}{\vartheta_{\uparrow ,
\downarrow}(x)} \, ;
\label{eq:34}
\end{equation}
the latter quantity may be interpreted as the transmission probability
corresponding to injection at the point $x$ toward the region containing
the maximum of the potential profile in the interval $[x',x'']$.
The ballistic velocity $v_{\uparrow , \downarrow}(x',x'';x)$ is given by
\begin{equation}
v_{\uparrow , \downarrow}(x',x'';x) = \frac{J^{l}_{\uparrow , \downarrow}
(x',x'';x)}{ n^{l}_{\uparrow , \downarrow}(x',x'';x)}
= \frac{2v_{e}}{C_{\uparrow , \downarrow}(x',x'';x)} \, .
\label{eq:35}
\end{equation}
It is not affected by spin relaxation, since the spin-flip mechanism is assumed
not to influence the kinematics of the electron motion.

From Eq.~(\ref{eq:31}), the continuation of {\em half} (see the remarks at the
beginning of Sec.~II.B) the densities (\ref{eq:17}) into the ballistic interval
now follows as
\begin{equation}
n^{l}_{\pm}(x',x'';x) = \smfrac{1}{2} \tilde{n}^{l}(x',x'';x)
[\alpha^{l}_{\pm }(x',x'';x) + P^m_{C}(x',x'';x)
\, \alpha^{l}_{\mp }(x',x'';x) ] \, ,
\label{eq:41}
\end{equation}
where
\begin{equation}
\tilde{n}^{l}(x',x'';x) \equiv \tilde{n}_{+}^{l}(x',x'';x) = \frac{N_c}{4}
D^m_{+}(x',x'';x) \, e^{\beta \tilde{\mu}(x')}
\label{eq:45}
\end{equation}
[see Eq.~(\ref{eq:03c})] is the total ballistic spin-relaxed  density,
and, in generalization of expression (\ref{eq:25b}) for $P^{m}(x',x'')$,
\begin{equation}
P^{m}_{C}(x',x'';x) = \frac{D_{-}^{m}(x',x'';x)}{D_{+}^{m}(x',x'';x)} \, .
\label{eq:41a}
\end{equation}
Extending condition (\ref{eq:10}) into the ballistic interval by requiring
\begin{equation}
n^{l}(x',x'';x) \equiv n^{l}_{+}(x',x'';x) = \tilde{n}^{l}(x',x'';x) \, ,
\label{eq:42aa}
\end{equation}
we find from Eqs.~(\ref{eq:41})
\begin{equation}
\alpha_{+}^{l}(x',x'';x) = 2 - P^{m}_{C}(x',x'';x) \alpha_{-}^{l}(x',x'';x)
\label{eq:41aa}
\end{equation}
[see Eq.~(\ref{eq:17a})], and hence for the ballistic spin-polarized density
inside the interval $[x',x'']$, in parallel to Eq.~(\ref{eq:17b}),
\begin{equation}
n^l_{-}(x',x'';x) = n^{l}(x',x'';x) \{ P^m_{C}(x',x'';x) + \smfrac{1}{2}
[Q_C^m(x',x'',x)]^2 \alpha_{-}^{l}(x',x'';x) \} \, ,
\label{eq:42a}
\end{equation}
where
\begin{equation}
[Q_C^m(x',x'';x)]^{2} \equiv  1 - [P_C^m(x',x'';x)]^2
= \frac{4D^m_{\uparrow}(x',x'';x)
D^m_{\downarrow}(x',x'';x)}{[D^m_+(x',x'';x)]^2}
\label{eq:42b}
\end{equation}
[see Eq.~(\ref{eq:27aa})].  Thus, we obtain for the ballistic equilibrium
spin-polarized density
\begin{equation}
\tilde{n}^{l}_{-}(x',x'';x)  =  \frac{N_c}{4}  D^m_-(x',x'';x) e^{\beta
\tilde{\mu}(x')}
\label{eq:47}
\end{equation}
[see Eq.~(\ref{eq:15c})], and for the ballistic off-equilibrium
spin-polarized density
\begin{equation}
\check{n}^{l}_{-}(x',x'';x) = \frac{N_c}{4} D^m(x',x'';x ) \, \,
e^{\beta \tilde{\mu}(x')} \alpha_{-}^{l}(x',x'';x)
\label{eq:49}
\end{equation}
[see Eq.~(\ref{eq:15d})], where
\begin{equation}
D^m(x',x'';x) =  \frac{2 D^m_{\uparrow}(x',x'';x ) \,
D^m_{\downarrow}(x',x'';x) }{D^m_+(x',x'';x)} = \smfrac{1}{2}
D_{+}^{m}(x',x'';x) [Q_{C}^{m}(x',x'';x)]^{2}  \, ,
\label{eq:49a}
\end{equation}
in parallel to Eq.~(\ref{eq:37e}).

We remark that for constant potential profiles, $\epsilon_{\uparrow ,
\downarrow}(x) \equiv\epsilon_{\uparrow , \downarrow}$, when
$C_{\uparrow , \downarrow}(x',x'';x) = 1$ so that $D_+^m(x',x'';x) =
\vartheta_+^m(x',x'')$, Eq.~(\ref{eq:45}) becomes $\tilde{n}^{l}(x',x'';x) =
\tilde{n}(x')/2 $. This reflects the fact that the left-hand side of this
relation refers to the density associated with the "half-sided" injected
current, while $\tilde{n}(x')$ is the total density at the point of thermal
equilibrium $x'$.

\subsection{Spatial behavior of the off-equilibrium spin-polarized current and
density}

The total ballistic current $J^l(x',x'')$ and density
$n^{l}(x',x'';x)$, as well as the ballistic equilibrium  spin-polarized
current $\tilde{J}^{l}_{-}(x',x'';x)$ and density
$\tilde{n}^{l}_{-}(x',x'';x)$, are determined by the quantities
$\vartheta^{m}_{\pm}(x',x'')$ and $D^{m}_{\pm}(x',x'';x)$ [which, in turn, are
completely determined by the potential profiles $\epsilon_{\uparrow ,
\downarrow}(x')$], and, most importantly, by the spin-relaxed chemical
potential $\tilde{\mu}(x')$, which is the only dynamical quantity appearing.
The latter is to be calculated according to the thermoballistic procedure
presented in Ref.~\onlinecite{lip05} (see below).

By contrast, the ballistic off-equilibrium  spin-polarized current
$\check{J}^{l}_{-}(x',x'';x)$ and density
$\check{n}^{l}_{-}(x',x'';x)$ contain, in addition, the common factor
${\alpha^{l}_{-}(x',x'';x)}$, whose spatial behavior ($x$-dependence)
is determined by the process of spin relaxation in the ballistic
interval $[x',x'']$. This process is governed by the balance equation
connecting the off-equilibrium ballistic spin-polarized current and
density (see Ref.~\onlinecite{lip05}),
\begin{equation}
\frac{d}{dx} \, \check{J}^{l}_{-}(x',x'';x) + \frac{2 v_{e}}{l_{s}} \,
\check{n}^{l}_{-}(x',x'';x) = 0 \, ;
\label{eq:50}
\end{equation}
here, $l_s = 2 v_e \tau_s$ is the (ballistic) spin relaxation length, which we
assume, as in the case of the effective electron mass $m^{\ast}$, to be
independent of position and of the external magnetic field.  The spin
relaxation time $\tau_{s}$ is given by $1/\tau_s = 1/\tau_{\uparrow
 \downarrow}
+ 1/\tau_{\downarrow  \uparrow}$, where $ 1/\tau_{\uparrow  \downarrow}$
$(1/\tau_{\downarrow  \uparrow})$ is the rate for spin-flip scattering from
spin-up (spin-down) to spin-down (spin-up) states.  Inserting expressions
(\ref{eq:37d}) and (\ref{eq:49}) in Eq.~(\ref{eq:50}), we now obtain a
first-order differential equation for $\alpha^{l}_{-}(x',x'';x)$,
\begin{equation}
\frac{d}{dx} \, \alpha^{l}_{-}(x',x'';x) + \frac{C(x',x'';x)}{l_{s}}
\,\alpha^{l}_{-}(x',x'';x) = 0 \, ,
\label{eq:51}
\end{equation}
where
\begin{equation}
C(x',x'';x) \equiv C_{\uparrow}(x',x'';x)\, C_{\downarrow}(x',x'';x)
\, \frac{\vartheta^{m}_{+}(x',x'')}{D^m_+(x',x'';x)} =
\frac{[Q^{m}_{C}(x',x'';x)]^{2}}{[Q^{m}(x',x'')]^{2}} \,
\frac{D^m_+(x',x'';x)}{\vartheta^m_+(x',x'')} \, .
\label{eq:52}
\end{equation}
The solution of Eq.~(\ref{eq:51}) obeying the initial condition
(\ref{eq:30}) is
\begin{equation}
\alpha_{-}^{l}(x',x'';x) = \alpha_{-}(x') \,
e^{-{\cal C}(x',x'';x',x)/l_{s}} \, ,
\label{eq:53}
\end{equation}
where
\begin{equation}
{\cal C}(x',x'';z_{1},z_{2}) = \int_{z_{<}}^{z_{>}} dz \,
C(x',x'';z) \, ,
\label{eq:54}
\end{equation}
with $z_{<} = \min (z_{1}, z_{2}), \; z_{>} = \max (z_{1},
z_{2})$.

For the ballistic off-equilibrium spin-polarized current, we then have from
Eq.~(\ref{eq:37d})
\begin{equation}
\check{J}^{l}_{-}(x',x'';x) = \frac{v_{e} N_{c}}{2} \,
\vartheta^{m}(x',x'') \,  A(x') \, e^{-{\cal C}(x',x'';x',x)/l_{s}} \, ,
\label{eq:54a}
\end{equation}
and for the ballistic off-equilibrium spin-polarized density
from Eq.~(\ref{eq:49}),
\begin{equation}
\check{n}^{l}_{-}(x',x'';x) = \frac{N_{c}}{4} D^m(x',x'';x )
A(x') e^{-{\cal C}(x',x'';x',x)/l_{s}} .
\label{eq:55a}
\end{equation}
Here, we have introduced the {\em spin transport function}
\begin{equation}
A(x') = e^{\beta \tilde{\mu}(x')} \, \alpha_{-}(x') = e^{\beta
\mu_{\uparrow}(x')} - e^{\beta \mu_{\downarrow}(x')} \, .
\label{eq:62}
\end{equation}
[Owing to the different definitions of $\alpha_{\uparrow , \downarrow}(x')$ in
the present work and in Ref.~\onlinecite{lip05}, the quantities $A(x')$ are not
the same in the two papers.] In the thermoballistic approach, this function
completely describes the dynamics of off-equilibrium spin-polarized transport.
In terms of the difference of the spin-resolved chemical potentials,
$\mu_{-}(x')$, which commonly has been used as the basic dynamic variable in
previous approaches to spin-polarized transport, it is expressed, owing to
Eq.~(\ref{eq:16a}), as
\begin{equation}
A(x') = \frac{2e^{\beta \tilde{\mu} (x')} \tanh \bm{(} \beta \mu_{-}
(x')/2 \bm{)} }{1 + P(x') \tanh \bm{(} \beta \mu_{-} (x')/2 \bm{)} } \,
.
\label{eq:62a}
\end{equation}
The procedure for the calculation of the spin transport function will
be outlined in Sec.~IV.C.

The spatial behavior of the off-equilibrium spin-polarized current and
density in the ballistic interval $[x',x'']$, i.e., their dependence
on the coordinate $x$, is determined by the factor
$e^{-{\cal C}(x',x'';x',x)/l_{s}}$. Hence, spin relaxation in
this interval departs from a purely exponential behavior unless the
potential profiles $\epsilon_{\uparrow , \downarrow}(x)$ are constant
over the interval.

\subsection{Net ballistic currents and joint ballistic densities}

So far, we have only considered injection from the left end-point,
$x'$, of the ballistic interval $[x',x'']$. The expressions for the
currents and densities obtained for that case are easily transcribed
to the analogous expressions for injection from the right end-point
$x''$, $J^{r}(x',x'')$, $\tilde{J}^{r}_{-}(x',x'';x)$,
$\check{J}^{r}_{-}(x',x'';x)$, $n^{r}(x',x'';x)$,
$\tilde{n}^{r}_{-}(x',x'';x)$, and $\check{n}^{r}_{-}(x',x'';x)$, by
replacing everywhere the superscript $l$ with $r$, and interchanging
the arguments $x'$ and $x''$ [owing to the symmetry of
$\epsilon^{m}_{\uparrow , \downarrow}(x',x'')$, the quantities
$\vartheta^{m}_{\pm}(x',x'')$, $\vartheta^{m}(x',x'')$,
$D^{m}_{\pm}(x',x'';x)$, and $D^{m}(x',x'';x)$ are symmetric under the
exchange of $x'$ and $x''$].

Now, combining the ballistic currents injected from the left and right, we
have for the (conserved) net {\em total} ballistic current inside the ballistic
interval, using Eq.~(\ref{eq:37b}),
\begin{equation}
J(x',x'') = J^{l}(x',x'') - J^{r}(x',x'')
= \frac{v_{e} N_{c}}{2} \, \vartheta^{m}_{+}(x',x'') \, [e^{\beta
\tilde{\mu}(x')}- e^{\beta \tilde{\mu}(x'')} ] \, .
\label{eq:57}
\end{equation}
Using Eq.~(\ref{eq:37c}), we find for the net ballistic {\em equilibrium}
{\em spin-polarized} current
\begin{eqnarray}
\tilde{J}_{-}(x',x'') &=& \tilde{J}^{l}_{-}(x',x'') - \tilde{J}^{r}_{-}(x',x'')
= J(x',x'') \, P^{m}(x',x'') \nonumber \\ &=& \frac{v_{e}
N_{c}}{2} \, \vartheta^{m}_{-}(x',x'') \, [e^{\beta \tilde{\mu}(x')}- e^{\beta
\tilde{\mu}(x'')} ] \, ,
\label{eq:60}
\end{eqnarray}
while for the net ballistic {\em off-equilibrium} {\em spin-polarized}
current, we have, using Eq.~(\ref{eq:54a}),
\begin{eqnarray}
\check{J}_{-}(x',x'';x) &=& \check{J}^{l}_{-}(x',x'';x) -
\check{J}^{r}_{-}(x',x'';x) \nonumber \\  &=& \frac{v_{e}
N_{c}}{2} \, \vartheta^{m}(x',x'') [ A(x') \, e^{-{\cal C}(x',x'';x',x)/l_{s}}
- A(x'') \, e^{-{\cal C}(x',x'';x,x'')/l_{s}} ] \, .
\label{eq:61}
\end{eqnarray}
Similarly, using Eq.~(\ref{eq:45}), we find for the joint {\em total}
ballistic  density associated with the net {\em total} ballistic
current (\ref{eq:57})
\begin{equation}
n(x',x'';x) = n^{l}(x',x'';x) + n^{r}(x',x'';x) =
\frac{N_c}{4} D^m_+(x',x'';x) \, [e^{\beta \tilde{\mu}(x')} + e^{\beta
\tilde{\mu}(x'')}] \, ,
\label{eq:57a}
\end{equation}
while for the joint ballistic {\em equilibrium spin-polarized} density
we obtain, using Eq.~(\ref{eq:47}),
\begin{eqnarray}
\tilde{n}_{-}(x',x'';x) &=& \tilde{n}^{l}_{-}(x',x'';x) +
\tilde{n}^{r}_{-}(x',x'';x) = n(x',x'';x) \, P^{m}_{C}(x',x'';x)
\nonumber \\ &=& \frac{N_{c}}{4} D^m_-(x',x'';x) \, [e^{\beta \tilde{\mu}(x')}
+ e^{\beta \tilde{\mu}(x'')}] \, .
\label{eq:60a}
\end{eqnarray}
Finally, for the joint ballistic {\em off-equilibrium spin-polarized density},
we have, using Eq.~(\ref{eq:55a}),
\begin{eqnarray}
\check{n}_{-}(x',x'';x) &=& \check{n}^{l}_{-}(x',x'';x) +
\check{n}^{r}_{-}(x',x'';x) \nonumber \\ &=& \frac{N_{c}}{4} \, D^m(x',x'';x )
[ A(x') \, e^{-{\cal C}(x',x'';x',x)/l_{s}} + A(x'') \,
e^{-{\cal C}(x',x'';x,x'')/l_{s}} ] .
\label{eq:61a}
\end{eqnarray}
It becomes apparent from the preceding results that the role of the
$x$-independent, local Boltzmann factors $\vartheta^m_{\uparrow ,
\downarrow}(x',x'')$ in the currents is taken over, in the densities, by the
$x$-dependent, nonlocal factors $D^m_{\uparrow , \downarrow}(x',x'';x)$.

\section{Thermoballistic spin-polarized transport}

The thermoballistic approach to semiclassical electron transport has been
developed in Ref.~\onlinecite{lip03}, and generalized in
Ref.~\onlinecite{lip05} to describe spin-polarized electron transport across a
spin-degenerate conduction band.  In this approach, the net {\em ballistic}
currents and joint {\em ballistic} densities defined in ballistic transport
intervals form the building blocks for the construction of the corresponding
{\em thermoballistic} currents and densities.  In order to deal with the case
of a spin-split conduction band, we have to modify and augment the formulation
of Ref.~\onlinecite{lip05}.  We construct the thermoballistic currents and
densities in terms of the net ballistic currents and joint ballistic densities
derived in Sec.~III.C, and outline the procedures for calculating the
equilibrium chemical potential and the spin transport function. From the
thermoballistic currents and densities, the corresponding equilibrium and
off-equilibrium spin polarizations are determined, and the magnetoresistance
is obtained.

\subsection{Thermoballistic currents and densities}

Following Refs.~\onlinecite{lip03} and \onlinecite{lip05}, we introduce, for
given position $x$ inside the semiconducting sample extending from $x_{1}$ to
$x_{2}$ (see Fig.~\ref{fig:1}), the set of ballistic intervals $[x',x'']$ that
enclose $x$, thus restricting $x'$ and $x''$ by $x_{1} \leq x' < x < x'' \leq
x_{2}$.  From the net ballistic currents and joint ballistic densities of
Eqs.~(\ref{eq:57})--(\ref{eq:61a}), here summarily represented by the symbol
${\cal F}(x',x'';x)$, the corresponding {\em thermoballistic} currents and
densities, ${\cal F}(x)$, are constructed by summing up the currents and
densities ${\cal F}(x',x'';x)$ over all ballistic intervals.  In the summation,
the ballistic intervals are weighted (assuming one-dimensional transport) with
the probability $e^{-|x'-x''|/l}$ that the electrons traverse the interval
without impurity or phonon scattering, multiplied by the probability $dx'/l$
($dx''/l$) of their being absorbed (or emitted) in an interval $dx'$ ($dx''$)
at an end-point $x'$ ($x''$).  At the ends of the semiconducting sample,
$x_{1,2}$, absorption and emission occur with unit probability.  Here, the
quantity $l$ is the momentum relaxation length (mean free path), which governs
the relative contribution of the ballistic and diffusive (``thermal'')
transport mechanisms (see Ref.~\onlinecite{lip03}).  Like the effective
electron mass $m^{\ast}$ and the ballistic spin relaxation length $l_{s}$, the
momentum relaxation length $l$ is assumed to be independent of position and of
the external magnetic field.

By this procedure, the thermoballistic currents and densities ${\cal F}(x)$ are
obtained in the form
\begin{eqnarray}
{\cal F}(x) &=& e^{-(x_{2}-x_{1})/l}  \; {\cal F}(x_{1},x_{2};x) +
\int_{x_{1}}^{x} \frac{dx'}{l} \, e^{-(x_{2}-x')/l}  \, {\cal
F}(x',x_{2};x) \nonumber \\  &+& \int_{x}^{x_{2}} \frac{dx''}{l} \,
e^{-(x''-x_{1})/l} \,{\cal F}(x_1, x'';x) + \int_{x_{1}}^{x}
\frac{dx'}{l}  \int_{x}^{x_2} \frac{dx''}{l} \, e^{-(x'' - x')/l}
\, {\cal F} (x',x'';x) \, .
\label{eq:64}
\end{eqnarray}
Thus, the total thermoballistic current ${\cal F}(x) \equiv
J(x)$ is given by Eq.~(\ref{eq:64}) with ${\cal F}(x',x'';x)$
replaced with $J(x',x'')$ of Eq.~(\ref{eq:57}).  Comparing
with Eq.~(2.12) of Ref.~\onlinecite{lip05}, we see that the current
$J(x)$ of the present work corresponds to the current $J(x)$
of Ref.~\onlinecite{lip05} if there one sets $E^0_c = 0$ and
substitutes for the weights $w_n(x',x'';l) \,[ n = 1,2,3]$ the
function
\begin{equation}
w_{+}(x',x'';l) = \smfrac{1}{2}\, e^{- |x' - x''|/l} \,
\vartheta^{m}_{+}(x',x'') \, ,
\label{eq:67}
\end{equation}
which in the limit of zero spin splitting, $\vartheta^{m}_{+}(x',x'')=2$,
reduces to the ($n$-independent) weight $w_n(x',x'';l)$ for
one-dimensional transport.

The equilibrium thermoballistic spin-polarized current
${\cal F}(x) \equiv \tilde{J}_{-}(x)$ follows from Eq.~(\ref{eq:64})
by identifying ${\cal F}(x',x'';x)$ with $\tilde{J}_{-}(x',x'')$ of
Eq.~(\ref{eq:60}). For the current $\tilde{J}_{-}(x)$, there is no
explicit counterpart in the formulation of Ref.~\onlinecite{lip05}, in
which the equilibrium part of the thermoballistic spin-polarized
current merely has been parameterized in terms of an arbitrary
constant $\tilde{\alpha}_{-}$ [see Eq.~(3.27) of Ref.~\onlinecite{lip05}].

The off-equilibrium thermoballistic spin-polarized current ${\cal
F}(x) \equiv \check{J}_{-}(x)$ is obtained from Eq.~(\ref{eq:64}) by
replacing ${\cal F}(x',x'';x)$ with $\check{J}_{-}(x',x'';x)$ of
Eq.~(\ref{eq:61}). The resulting current $\check{J}_{-}(x)$ is seen to
correspond to the current given by Eq.~(3.28) of
Ref.~\onlinecite{lip05} if the function
\begin{equation}
\check{w}(x',x'';l) = \smfrac{1}{2}\, e^{- |x' - x''|/l} \,
\vartheta^{m}(x',x'')
\label{eq:68}
\end{equation}
is substituted for the weights $w_n(x',x'';l)$. Here again, in the
limit of zero spin splitting, $\vartheta^{m}(x',x'')=1$, the function
$\check{w}(x',x'';l)$ agrees with the weight $w_n(x',x'';l)$ for
one-dimensional transport.

We note that the total and equilibrium spin-polarized currents, $J(x)$ and
$\tilde{J}_{-}(x)$, both are {\em linear} functionals of the factors
$\vartheta_{\pm}^{m}(x',x'')$, and hence of the Boltzmann factors corresponding
to the maxima of the potential profiles $\epsilon_{\uparrow , \downarrow}(x)$
in the ballistic interval $[x',x'']$.  The off-equilibrium current
$\check{J}_-(x)$, on the other hand, is, via its dependence on the factor
$\vartheta^{m}(x',x'')$, a {\em nonlinear} functional of the Boltzmann factors.

The thermoballistic densities $n(x)$, $\tilde{n}_{-}(x)$, and
$\check{n}_{-}(x)$ corresponding to the thermoballistic currents
$J(x)$, $\tilde{J}_{-}(x)$, and $\check{J}_{-}(x)$, respectively, are
evidently obtained by substituting the joint ballistic densities of
Eqs.~(\ref{eq:57a})--(\ref{eq:61a}) for ${\cal F}(x',x'';x)$ in
Eq.~(\ref{eq:64}).  Comparing the thermoballistic densities of the present work
to the densities given by Eqs.~(2.17), (3.29), and (3.30) of
Ref.~\onlinecite{lip05}, we find that the remarks made above in conjunction
with the thermoballistic currents also apply here if for the weights
given by Eq.~(2.55) of Ref.~\onlinecite{lip05} the functions $w_{+}(x',x'';l)$
and $\check{w}(x',x'';l)$ of Eq.~(\ref{eq:67}) and (\ref{eq:68}), respectively,
are used, with $\vartheta^{m}_{+}(x',x'')$ and $\vartheta^{m}(x',x'')$ replaced
with $D^{m}_{+}(x',x'';x)$ and $D^{m}(x',x'';x)$, respectively.

\subsection{Spin-relaxed chemical potential and equilibrium spin polarizations}

The net total ballistic and ballistic equilibrium spin-polarized currents,
Eqs.~(\ref{eq:57}) and (\ref{eq:60}), respectively, as well as the joint total
ballistic and ballistic equilibrium spin-polarized densities,
Eqs.~(\ref{eq:57a}) and (\ref{eq:60a}), respectively, and hence also the
corresponding thermoballistic currents and densities, are dynamically
determined by the spin-relaxed chemical potential $\tilde{\mu}(x)$.  The
procedure for calculating the latter quantity is set up \cite{lip05} by
employing the conditions
\begin{equation}
\frac{1}{x_{2} - x_{1}} \, \int_{x_{1}}^{x_{2}} dx \, J(x) = J
\, ,
\label{eq:71}
\end{equation}
where $J$ is the total physical current, and
\begin{equation}
J(x^{+}_{1}) = J(x^{-}_{2})
\label{eq:72}
\end{equation}
for the total thermoballistic current $J(x)$ given by Eq.~(\ref{eq:64}) with
${\cal F}(x) \equiv J(x)$.

We now rewrite $J(x)$, with the help of Eq.~(\ref{eq:57}), in
terms of the (spin-relaxed) ``resistance function'' \cite{lip03,lip05}
\begin{equation}
\chi(x) = \frac{v_e N_{c}}{J} \, [e^{\beta \tilde{\mu}(x_1)}-
e^{\beta \tilde{\mu}(x)}] \, .
\label{eq:73}
\end{equation}
Inserting the result in Eq.~(\ref{eq:71}) and replacing the parameter $x_{2}$
with the variable $x$, we obtain a linear, inhomogeneous, Volterra-type
integral equation for $\chi(x)$,
\begin{equation}
\frac{x-x_{1}}{l} - f_{+}(x;x_{1};l) \chi(x) + \int_{x_{1}}^{x}
\frac{dx'}{l} \, K_{+}(x,x';x_{1};l) \, \chi(x') = 0 \, ,
\label{eq:73a}
\end{equation}
where
\begin{equation}
f_{+}(x;x_{1};l) = u_{+}(x_{1},x;l) + \int_{x_{1}}^{x} \frac{dx'}{l}
\, u_{+}(x',x;l)
\label{eq:74}
\end{equation}
and
\begin{equation}
K_{+}(x,x';x_{1};l) = u_{+}(x',x;l) - u_{+}(x_{1},x';l)  +
\int_{x_{1}}^{x} \frac{dx''}{l} \, u_{+}(x',x'';l) \, ,
\label{eq:75}
\end{equation}
with
\begin{equation}
u_{+}(x',x'';l)  = \frac{x''-x'}{l} \, w_{+}(x',x'';l) \, ,
\label{eq:76}
\end{equation}
where $w_{+}(x',x'';l)$ is given by Eq.~(\ref{eq:67}). At $x=x_{1}$, the
resistance function $\chi(x)$ is discontinuous,
\begin{equation}
\chi(x_{1}) = 0 \, ,
\label{eq:76abcd}
\end{equation}
\begin{equation}
\chi(x_{1}^{+}) = \frac{2}{\vartheta^{m}_{+}(x_{1},x_{1}^{+})} =
\frac{2}{\vartheta^{m}_{+}(x_{1},x_{1})}
\label{eq:76abc}
\end{equation}
(note that we assume the potential profiles $\epsilon_{\uparrow ,
\downarrow}(x) $ to be continuous in the interval $[x_{1}, x_{2}]$).  The
function $f_{+}(x;x_{1};l)$ and the kernel $K_{+}(x,x';x_{1};l)$ are seen to
depend on the profiles $\epsilon_{\uparrow , \downarrow}(x)$ solely via the
quantity $\vartheta^{m}_{+}(x',x'')$.  The solution of Eq.~(\ref{eq:73a}) is
obtained, in general, by numerical propagation.

In order to construct a unique chemical potential $\tilde{\mu}(x)$, following
Ref.~\onlinecite{lip05}, we have to solve Eq.~(\ref{eq:73a}) twice:\ first,
with the original profiles $\epsilon_{\uparrow , \downarrow}(x)$ to obtain the
solution $\chi_{1}(x)$, and second, with the spatially reversed profiles
$\epsilon^{*}_{\uparrow , \downarrow}(x) = \epsilon_{\uparrow
, \downarrow}(x_{1}
+ x_{2} - x)$ to obtain the solution $\chi_{1}^{*}(x)$. Defining
\begin{equation}
\chi_{2}(x) = \chi_{1}^{*}(x_{1} + x_{2} - x) \, ,
\label{eq:76a}
\end{equation}
we introduce the function
\begin{equation}
\chi_{-}(x) = \hat{a}_{1} [ \chi_{1}(x) - \smfrac{1}{2} \chi_{1}(x_{2}) ] -
\hat{a}_{2} [ \chi_{2}(x) - \frac{1}{2} \chi_{2}(x_{1}) ]
\label{eq:77}
\end{equation}
and the constant
\begin{equation}
\chi = \hat{a}_{1} \chi_{1}(x_{2}) + \hat{a}_{2} \chi_{2}(x_{1}) \, .
\label{eq:78}
\end{equation}
The coefficients $\hat{a}_{1,2}$ are given \cite{lip05} by
\begin{equation}
\hat{a}_{1,2} = \frac{a_{1,2}}{a} \, ,
\label{eq:78a}
\end{equation}
with
\begin{equation}
a_{1} = \int_{x_{1}}^{x_{2}} \frac{dx'}{l} \, \{ w_{+}(x_{1}, x';l) \,
[ \chi_{2}(x') - \chi_{2}(x_{1})] +
w_{+}(x',x_{2};l) \chi_{2}(x') \} \label{eq:78b} \, ,
\end{equation}
\begin{equation}
a_{2} = \int_{x_{1}}^{x_{2}} \frac{dx'}{l} \, \{ w_{+}(x_{1}, x';l) \,
\chi_{1}(x') + w_{+}(x', x_{2};l) \, [
\chi_{1}(x') - \chi_{1}(x_{2})] \} \, ,
\label{eq:78c}
\end{equation}
and $a = a_{1} + a_{2}$, so that $\hat{a}_{1} + \hat{a}_{2} =
1$. For symmetric potential profiles, $\epsilon^{*}_{\uparrow ,
\downarrow}(x) = \epsilon_{\uparrow , \downarrow}(x)$, we have
$\chi_{2}(x) = \chi_{1}(x_{1} + x_{2} -x)$, and hence $\hat{a}_{1} =
\hat{a}_{2} = \frac{1}{2}$.

In terms of $\chi_{-}(x)$ and $\chi$, the {\em thermoballistic} spin-relaxed
chemical potential $\tilde{\mu}(x)$ is given \cite{lip05} by
\begin{equation}
e^{\beta \tilde{\mu}(x)} = \eta_{+} - 2 \, \frac{\chi_{-}(x)}{\chi} \,
\eta_{-}
\label{eq:79}
\end{equation}
$(x_{1} \leq x \leq x_{2})$, where
\begin{equation}
\eta_{\pm} = \smfrac{1}{2} ( e^{\beta \tilde{\mu}_{1}} \pm e^{\beta
\tilde{\mu}_{2}})
\label{eq:80}
\end{equation}
and $\tilde{\mu}_{1,2} = \tilde{\mu}(x_{1,2})$.  The chemical potential
$\tilde{\mu}(x)$ thus obtained in the thermoballistic approach characterizes
the thermal-equilibrium attribute of the thermoballistic system, which
co-exists with its ballistic attribute.  It immediately leads, via
Eqs.~(\ref{eq:03c}) and (\ref{eq:15c}), to the total spin-relaxed density
$n(x)$ and the equilibrium spin-polarized density $\tilde{n}_-(x)$,
respectively.

By analogy with Eq.~(2.44) of Ref.~\onlinecite{lip05}, we introduce the
(dimensionless) reduced resistance
\begin{equation}
\tilde{\chi} = \smfrac{1}{2} \vartheta^{m}_{+}(x_{1},x_{2}) \, \chi \, ,
\label{eq:80lala}
\end{equation}
which fixes, for given total current $J$, via the current-voltage
characteristic
\begin{equation}
J = \smfrac{1}{2} v_{e} N_{c} \, \vartheta^{m}_{+}(x_{1}, x_{2}) \, e^{\beta
\tilde{\mu}_{1}} \, \frac{1}{\tilde{\chi}} \, ( 1 - e^{-\beta e \tilde{V}})
\label{eq:80a}
\end{equation}
[see Eq.~(2.42) of Ref.~\onlinecite{lip05}] the "spin-relaxed" voltage bias
\begin{equation}
\tilde{V} = \frac{\tilde{\mu}_{1} - \tilde{\mu}_{2}}{e}
\label{eq:80b}
\end{equation}
between the contacts bordering the semiconducting sample.  In the zero-bias
limit, $\tilde{\chi}$ determines, via its dependence on the Zeeman splitting
$\Delta(x)$, the magnetic-field-dependent {\em equilibrium} resistance
\begin{equation}
\tilde{R} = \left. \frac{\tilde{V}}{eJ} \right|_{J \rightarrow 0} =
\frac{\tilde{\chi} e^{- \beta E_{c}(x_{1})}}{\beta e^{2} v_{e}
n_{1}^{(0)}} \, \frac{2}{\vartheta^{m}_{+}(x_{1}, x_{2})} \, ,
\label{eq:80c}
\end{equation}
where $n_{1}^{(0)} = N_{c} \, e^{- \beta [E_{c}(x_{1}) -\tilde{\mu}_{1}]}$ is
the total (spin-relaxed) electron density at $x_{1}$ for
zero external magnetic field, i.e., for $\Delta (x) \equiv 0$. [Note that,
owing to its definition in terms of a current {\em density}, $\tilde{R}$ has
the dimension of an {\em interface} resistance.] It is notationally convenient,
and physically meaningful, to introduce, instead of the density $n_{1}^{(0)}$,
the Sharvin interface conductance \cite{sha65}
\begin{equation}
{\cal G}_{1}^{(0)} = \beta e^{2} v_{e} n_{1}^{(0)} \, ,
\label{eq:66zog}
\end{equation}
so that
\begin{equation}
\tilde{R} = \frac{\tilde{\chi} e^{- \beta E_{c}(x_{1})}}{{\cal G}_{1}^{(0)}} \,
\frac{2}{\vartheta^{m}_{+}(x_{1}, x_{2})} \, .
\label{eq:80cia}
\end{equation}
Adding to the spin-relaxed bias $\tilde{V}$ the off-equilibrium bias, one
obtains the {\em magnetoresistance} proper (see Sec.~IV.D).

The {\em total thermoballistic current} $J(x)$ is obtained in terms of
$\tilde{\mu}(x)$ or, equivalently, $\chi_{-}(x)$ and $\tilde{\chi}$ by
substituting expression (\ref{eq:57}) in Eq.~(\ref{eq:64}) with ${\cal F}(x)
\equiv J(x)$.  In an analogous manner, the total thermoballistic
density $n(x)$ associated with the total thermoballistic current is
obtained by referring to Eqs.~(\ref{eq:57a}) and (\ref{eq:64}).

We define the equilibrium current spin polarization $\tilde{P}_{J}(x)$
and the equilibrium density spin polarization $\tilde{P}_{n}(x)$ in
terms of the thermoballistic currents and densities, respectively, as
\begin{equation}
\tilde{P}_{J}(x) = \frac{\tilde{J}_{-}(x)}{J(x)}
\label{eq:65}
\end{equation}
and
\begin{equation}
\tilde{P}_{n}(x) = \frac{\tilde{n}_{-}(x)}{n(x)} \; .
\label{eq:66}
\end{equation}
The thermoballistic currents and densities are, in general, not equal
to the corresponding physical currents and densities. However, the
{\em relative} spin content is the same in both, so that the
``thermoballistic'' ratio in Eqs.~(\ref{eq:65}) and (\ref{eq:66}),
respectively, is equal to the corresponding ``physical'' ratio.

In view of Eq.~(\ref{eq:79}) and the first of Eqs.~(\ref{eq:80a}),
the term in brackets on the right-hand side of Eq.~(\ref{eq:57}) for
the net total ballistic current $J(x',x'')$ is proportional to
the total physical current $J$, since the function $\chi_-(x)$ is
independent of $J$. From expression (\ref{eq:64}), the total
thermoballistic current $J(x)$ is then again found to be
proportional to $J$, and the same holds for the equilibrium
thermoballistic spin-polarized current $\tilde{J}_{-}(x)$ [see
Eqs.~(\ref{eq:60}) and (\ref{eq:64})]. The equilibrium current spin
polarization $\tilde{P}_{J}(x)$ is therefore independent of $J$
(it is an intrinsic property determined exclusively by the spin
splitting of the conduction band inside the semiconductor). The
equilibrium density spin polarization $\tilde{P}_{n}(x)$, on the other
hand, is found to depend, in general, on the bias $\tilde{V}$, i.e., on $J$
[see Eq.~(2.53) of Ref.~\onlinecite{lip05}].

\subsection{Spin transport function and off-equilibrium spin polarizations}

The off-equilibrium thermoballistic spin-polarized current
$\check{J}_{-}(x)$ is obtained by substituting expression (\ref{eq:61}) in
Eq.~(\ref{eq:64}) with ${\cal F}(x) \equiv \check{J}_{-}(x)$, and
analogously for the corresponding density (\ref{eq:61a}).  Following
Ref.~\onlinecite{lip05}, we invoke the spin balance equation
connecting these two quantities and arrive at an integral equation for
the spin transport function $A(x)$ of the form [see
Eqs.~(3.35)--(3.37) of Ref.~\onlinecite{lip05}]
\begin{equation}
\check{\cal W}(x_{1},x;l,l_{s}) \, A_{1} + \check{\cal
W}(x,x_{2};l,l_{s}) \, A_{2} - \check{W}(x;x_{1},x_{2};l) \, A(x)
+ \int_{x_{1}}^{x_{2}} \frac{dx'}{l} \, \check{\cal W}(x',x;l,l_{s}) \, A(x')
= 0 \; ,
\label{eq:67ax}
\end{equation}
where
\begin{equation}
\check{\cal W}(x',x'';l,l_{s}) = \check{w}(x',x'';l) \, e^{- {\cal C}
(x',x'';x',x'')/l_{s}}
\label{eq:68b}
\end{equation}
and
\begin{equation}
\check{W}(x;x_{1},x_{2};l) = \check{w}(x_{1},x;l) + \check{w} (x,x_{2};l)
+ \int_{x_{1}}^{x_{2}}\frac{dx'}{l}\check{w}(x',x;l) \; .
\label{eq:69x}
\end{equation}
Here, $A_{1,2} = A(x_{1,2})$ are the values of the spin transport function at
the contact side of the interfaces with the left and right contacts,
respectively; they are determined by the chemical potentials
$\mu_{\uparrow , \downarrow}(x_{1,2})$ at these positions [see
Eq.~(\ref{eq:62})].
The exponent ${\cal C}(x',x'';x',x'')$ is given by Eq.~(\ref{eq:54}), and
$\check{w}(x',x'';l)$ by Eq.~(\ref{eq:68}).  Equation (\ref{eq:67ax}) is a
linear, inhomogeneous, Fredholm-type integral equation for the spin transport
function $A(x)$.  The corresponding homogeneous equation is solved by $A(x)
\equiv 0$ only, so that the solution $A(x)$ of Eq.~(\ref{eq:67ax}) for $x_{1} <
x < x_{2}$ is linear and homogeneous in $A_{1}$ and $A_{2}$.  The function
$A(x)$ is not, in general, continuous at the interfaces with the contacts,
$A(x_1^+) \neq A_{1}, A(x_2^-) \neq A(x_{2})$, as a consequence of analogous
discontinuities of the spin-resolved chemical potentials $\mu_{\uparrow ,
\downarrow}(x)$ [``Sharvin effect''].  In general, Eq.~(\ref{eq:67ax}) can be
solved numerically by using matrix methods after discretization.

The off-equilibrium current spin polarization is given by
\begin{equation}
\check{P}_{J}(x) =
\frac{\check{J}_{-}(x)}{J(x)} \, ,
\label{eq:65ax}
\end{equation}
and similarly, the off-equilibrium density spin polarization, by
\begin{equation}
\check{P}_{n}(x) =
\frac{\check{n}_{-}(x)}{n(x)} \, .
\label{eq:66a}
\end{equation}
Here again, as in the case of the spin-relaxed polarizations, we calculate the
polarizations in terms of thermoballistic currents and densities, respectively.

These off-equilibrium polarizations are determined, via Eq.~(\ref{eq:64}), by
the current $\check{J}_{-}(x',x'';x)$ and density $\check{n}_{-}(x',x'';x)$,
Eqs.~(\ref{eq:61}) and (\ref{eq:61a}), respectively, which are linearly
connected with the spin transport function $A(x)$, and therefore, linear in the
boundary values $A_{1,2}$,
\begin{equation}
\frac{\check{J}_{-}(x)}{v_{e} N_{c}} = F_{1}(x) \, A_{1} + F_{2}(x) \,
A_{2} \, ,
\label{eq:91x}
\end{equation}
with ``formfactors''  $F_{1}(x)$ and $F_{2}(x)$.

It remains to determine the quantities $A_{1,2}$, or rather
$\alpha_{1,2} = \alpha_-(x_{1,2})$ [see Eq.~(\ref{eq:62})].  We do
this by making use of the continuity of the total current spin polarization
\begin{equation}
P_J(x) = \tilde{P}_J(x)  + \check{P}_J(x) \, ,
\label{eq:91xx}
\end{equation}
closely following Sec.~IV of Ref.~\onlinecite{lip05}. At the
interfaces with the semiconductor at $x = x_{1}$ and $x = x_{2}$,
respectively, the current polarizations $P_J(x_{1,2})$ in the left and
right (semi-infinite) external {\em contacts}, which we take as nonmagnetic
metals, are found as
\begin{equation}
P_{J}(x_{1,2}) = \mp \frac{G_{1,2}}{2 \beta e^{2} J} \, \ln \left(\frac{1 +
\frac{1}{2} [1 - P_{1,2}] \alpha_{1,2}}{1- \frac{1}{2} [1 + P_{1,2}]
\alpha_{1,2}} \right) \, ;
\label{eq:66b}
\end{equation}
here,
\begin{equation}
G_{1,2} = \frac{\sigma_{1,2}}{L_s^{(1,2)}}  \, ,
\label{eq:66bbb}
\end{equation}
where $\sigma_{1,2}$ are the conductivities of the two contacts,
$L^{(1,2)}_{s}$ the spin diffusion lengths, and $P_{1,2} = P(x_{1,2})$.  We
note that Eq.~(\ref{eq:66b}) differs from Eqs.~(4.7) and (4.8) of
Ref.~\onlinecite{lip05} with respect to the role of the polarizations
$P_{1,2}$:\ in
the former equation, $P_{1,2}$ refer to the paramagnetic semiconductor, while
in the latter equations they refer to the ferromagnetic leads (note also that
the normalization of the quantities $\alpha_{1,2}$ used here differs from that
used in Ref.~\onlinecite{lip05}).  The corresponding values of the polarization
in the {\em semiconductor} are given by
\begin{equation}
P_J(x_{1,2}) = \tilde{P}_J(x_{1,2}) \  +
\frac{v_eN_c}{\kappa_{1,2} J} [F_{1}(x_{1,2}) \, A_{1} + F_{2}(x_{1,2}) \,
A_{2}] \, ,
\label{eq:66c}
\end{equation}
with $\kappa_{1,2} = J(x_{1,2})/J$. Equating the right-hand
sides of Eqs.~(\ref{eq:66b}) and (\ref{eq:66c}), we obtain a pair of
nonlinear equations which determine the quantities $\alpha_{1,2} =
A_{1,2} e^{-\beta \tilde{\mu}(x_{1,2})}$, and thereby, the
off-equilibrium current spin polarization $\check{P}_J(x)$, in terms of
the equilibrium polarization $\tilde{P}_J(x)$, i.e., of the Zeeman
splitting $\Delta(x)$, and the material parameters of the metal
contacts and the semiconductor.

In the zero-bias limit, we have $|\alpha_{1,2}| \ll 1$, and
the pair of equations for $\alpha_{1,2}$ become linear,
\begin{equation}
\mp \frac{G_{1,2}}{ 2 \beta e^{2}} \, \alpha_{1,2} -  \frac{v_e
\tilde{n}(x_{1,2})}{\kappa_{1,2}} \, [F_1(x_{1,2}) \alpha_{1} + F_2(x_{1,2})
\alpha_{2}] = \tilde{P}_J(x_{1,2})J \, .
\label{eq:66e}
\end{equation}
As solutions of Eq.~(\ref{eq:66e}), the quantities $\alpha_{1,2}$ are
proportional to the current $J$, so that the current spin polarization
$P_{J}(x)$ is independent of $J$.

Summing up, we obtain the current spin polarization along the entire
heterostructure, $P_{J}(x)$, as follows. In the nonmagnetic contacts, it
decays exponentially away from the interfaces at $x_{1,2}$ with decay
length $L_{s}^{(1,2)}$,
\begin{equation}
P_{J}(x) = \mp \frac{G_{1,2}}{2 e^{2} J} \, \mu_{-}(x_{1,2}) \,
e^{-|x_{1,2} - x|/L_{s}^{(1,2)}}
\label{eq:66ikea}
\end{equation}
[see Eqs.~(4.3) and (4.5) of Ref.~\onlinecite{lip05} with $P_{1,2} = 0$].  In
the semiconductor, it is given by expression (\ref{eq:91xx}), with the
formfactors $F_{1,2}(x)$ determined by the solution of the integral equation
(\ref{eq:67ax}).  The explicit form of the latter depends on the structure of
the semiconductor.

\subsection{Magnetoresistance}

Once the quantities $\alpha_{1,2}$ are determined, we can obtain the
magnetoresistance $R = [V/eJ]_{J \rightarrow 0}$, where $V$ is the voltage
bias related to the mean chemical potentials $\bar{\mu}_{1,2} =
\bar{\mu}(x_{1,2})$ at the contact side of the contact-semiconductor
interfaces,
\begin{equation}
V = \frac{1}{e}(\bar{\mu}_{1} - \bar{\mu}_{2})
\label{eq:66f}
\end{equation}
(here, we assume the metal contacts to have infinitely high conductivity, so
that there is no voltage drop inside the contacts).  From  Eq.~(\ref{eq:16aa}),
we have
\begin{equation}
\bar{\mu}_{1,2} = \tilde{\mu}_{1,2} + \frac{1}{2\beta} \, \ln \bm{(}
\alpha_{\uparrow}(x_{1,2}) \alpha_{\downarrow}(x_{1,2}) \bm{)} \, ,
\label{eq:66gg}
\end{equation}
which in the zero-bias case reduces to
\begin{equation}
\bar{\mu}_{1,2} = \tilde{\mu}_{1,2} - \frac{1}{2\beta} \, P_{1,2}
\alpha_{1,2} \, .
\label{eq:66h}
\end{equation}
Subtracting the two equations (\ref{eq:66h}) from one another, we
obtain the magnetoresistance as
\begin{equation}
R = \tilde{R} + \check{R} \, ,
\label{eq:66i}
\end{equation}
where the equilibrium contribution $\tilde{R}$ is given by
Eq.~(\ref{eq:80cia}), and
\begin{equation}
\check{R} = - \frac{1}{2\beta e ^{2} J} ( P_{1} \alpha_{1} - P_{2}
\alpha_{2} )
\label{eq:66ii}
\end{equation}
is the off-equilibrium contribution.

\section{DMS/NMS/DMS heterostructures:\ Theory}

As an illustrative example, we now consider the thermoballistic description of
spin-polarized electron transport in a heterostructure formed of an NMS layer
sandwiched between two DMS layers.  As a structure of this kind is composed
entirely of semiconducting material, one should ideally treat it as a {\em
single} sample, so that the ballistic intervals $[x',x'']$ may enclose one or
both of the DMS/NMS interfaces.  In this case, one has to consider carefully
the effect of the interfaces on the electron motion.  The interfaces are
characterized by (i) abrupt changes in the material parameters, reflecting
changes in impurity and phonon scattering and in magnetic scattering, (ii) high
structural disorder in their vicinity, and (iii) an abrupt change in the
electrostatic potential due to the change in spin splitting.  Taking into
account the combined effect of these features in a consistent way is beyond the
scope of the present work.

Since, here, we place emphasis on the general formulation of the theory, we
adopt a pragmatic point of view:\ we assume the different layers in a
DMS/NMS/DMS heterostructure to be homogeneous and require the interfaces to act
as fixed points of thermal equilibrium.  That is, we apply the thermoballistic
description separately to the different layers (thus dealing with conduction
band potentials and spin splittings that are, in general, {\em discontinuous}
at the interfaces), evaluating for each layer the spin transport function and
the current spin polarization for a {\em homogeneous} semiconductor, and
subsequently connect these functions across the interfaces to obtain the
position dependence of the spin polarization as well as the magnetoresistance.
In a previous publication \cite{lip07}, we have applied this concept within a
heuristic approach.

\subsection{Homogeneous semiconductor}

Inside a homogeneous semiconducting layer, at zero bias, we have a flat
spin-independent potential profile, $E_{c}(x) \equiv E_{c}$, and a constant
spin splitting, $\Delta(x) \equiv \Delta$, so that the profiles
$\epsilon_{\uparrow , \downarrow}(x)$ are given by
\begin{equation}
\epsilon_{\uparrow , \downarrow}(x) \equiv  E_{c} \pm \Delta /2
\label{eq:66iii}
\end{equation}
$(x_{1} \leq x \leq x_{2})$.  For the functions $ \vartheta^{m}_{+}(x',x'')$
and $\vartheta^{m}_{-}(x',x'')$, respectively, we then have from
Eq.~(\ref{eq:03b})
\begin{equation}
\vartheta^{m}_{+}(x',x'') \equiv 2 e^{-\beta E_{c}} \cosh(\beta \Delta /2) =
\frac{2}{Q} e^{-\beta E_{c}}
\label{eq:66x}
\end{equation}
and
\begin{equation}
\vartheta^{m}_{-}(x',x'') \equiv - 2 e^{-\beta E_{c}} \sinh(\beta \Delta /2)
=  \frac{2P }{Q} e^{-\beta E_{c}}  \, ,
\label{eq:66y}
\end{equation}
where $Q = (1 - P^{2})^{1/2}$, and $P = - \tanh (\beta \Delta /2)$ is the
static spin polarization.  Now, using Eqs.~(\ref{eq:67}) and (\ref{eq:66x}) in
Eq.~(\ref{eq:76}), we easily see from Eqs.~(\ref{eq:73a})--(\ref{eq:75}) that
the resistance functions $\chi_{1}^{h}(x)$ and $\chi_{2}^{h}(x)$ for the
homogeneous semiconductor are given by the corresponding functions for
$E_{c}(x) \equiv \Delta(x) \equiv 0$ [see Eq.~(2.68) of
Ref.~\onlinecite{lip05}], multiplied by $e^{\beta E_{c}} Q$.  Hence, the
function $\chi_{-}^{h}(x)/\chi_{h}$ is independent of $E_{c}$ and $\Delta$.
Equations (\ref{eq:79}) and (\ref{eq:80}) show that, for given
$\tilde{\mu}_{1,2}$, the spin-relaxed chemical potential
$\tilde{\mu}_{h}(x)$ is also independent of $E_{c}$ and $\Delta$.  Using
Eqs.~(\ref{eq:57}) and (\ref{eq:66x}) in Eq.~(\ref{eq:64}), we then find the
total thermoballistic current in the semiconductor as
\begin{equation}
J_{h}(x) \equiv J_{h}  = \frac{e^{-\beta E_{c}}}{Q} \,
J^{(0)}_{h} \, ,
\label{eq:66z}
\end{equation}
where $J^{(0)}_{h}$ is the (conserved) total physical current for $E_{c} =
\Delta = 0$. Similarly, using Eqs.~(\ref{eq:60}) and (\ref{eq:66y}) in
Eq.~(\ref{eq:64}), we have for the thermoballistic equilibrium
spin-polarized current
\begin{equation}
\tilde{J}^{h}_{-}(x) \equiv \tilde{J}^{h}_{-} = e^{-\beta E_{c}} \, \frac{P}{Q}
\, J^{(0)}_{h} \, ,
\label{eq:66za}
\end{equation}
and hence for the zero-bias equilibrium current spin polarization of the
homogeneous semiconductor
\begin{equation}
\tilde{P}^{h}_{J}(x) = \frac{\tilde{J}^{h}_{-}(x)}{J_{h}(x)} \equiv
\frac{\tilde{J}^{h}_{-}}{J_{h}}= P \, ,
\label{eq:66zb}
\end{equation}
thereby retrieving the static spin polarization, as was to be expected.

As to the off-equilibrium current spin polarization, we have from
Eq.~(\ref{eq:37e})
\begin{equation}
\vartheta^{m}(x',x'') \equiv \frac{e^{- \beta E_{c}}}{\cosh(\beta \Delta /2)} =
Q e^{- \beta E_{c}} \, ;
\label{eq:66ijk}
\end{equation}
this factor drops out from the integral equation (\ref{eq:67ax}), and as its
solution, the spin transport function $A(x)$ is seen to be equal to the
function for $E_{c} = \Delta = 0$, given by Eq.~(3.56) of
Ref.~\onlinecite{lip05},
\begin{equation}
A(x) = C_{1} e^{- (x - x_{1})/L} + C_{2} e^{- (x_{2} - x)/L} ,
\label{eq:90a}
\end{equation}
where the generalized spin diffusion length $L$ is given by
\begin{equation}
L = \sqrt{\bar{l} l_{s}} = \frac{L_{s}}{\sqrt{1 + l/l_{s}}} \, ,
\label{eq:90b}
\end{equation}
with
\begin{equation}
\frac{1}{\bar{l}} = \frac{1}{l} +
\frac{1}{l_{s}} \, ,
\label{eq:90bb}
\end{equation}
and
\begin{equation}
L_{s} = \sqrt{l l_{s}}
\label{eq:90bbb}
\end{equation}
is the spin diffusion length proper. The coefficients $C_{1,2}$ are linearly
connected with the boundary values $A_{1,2}$ [see Eqs.~(3.57) and (3.58) of
Ref.~\onlinecite{lip05}].

Using Eq.~(\ref{eq:61}) with Eq.~(\ref{eq:66ijk}) in Eq.~(\ref{eq:64}), we find
for the thermoballistic off-equilibrium spin-polarized current
\begin{equation}
\check{J}^{h}_{-}(x) = Q e^{- \beta E_{c}} \, \check{J}^{h(0)}_{-}(x) \, ,
\label{eq:66zc}
\end{equation}
where $\check{J}^{h(0)}_{-}(x)$ is given by
\begin{equation}
\check{J}^{h(0)}_{-}(x) = - v_{e} N_{c} \bar{l} \, \frac{dA(x)}{dx}
\label{eq:66zf}
\end{equation}
[see Eq.~(4.11) of Ref.~\onlinecite{lip05}; the factor 2 appearing in the
right-hand side of the latter equation reflects the fact that the normalization
of $A(x)$ used there differs from that used in the present paper].  Combining
expressions (\ref{eq:66zc}) and (\ref{eq:66z}), we obtain the
zero-bias off-equilibrium current spin polarization of the homogeneous
semiconductor as
\begin{equation}
\check{P}^{h}_{J}(x) = \frac{\check{J}^{h}_{-}(x)}{J_{h}(x)} = Q^{2}
\, \frac{\check{J}^{h(0)}_{-}(x) }{J^{(0)}_{h} } \, ,
\label{eq:66zd}
\end{equation}
and hence
\begin{equation}
P^{h}_{J}(x) = \tilde{P}^{h}_{J}(x) + \check{P}^{h}_{J}(x) = P + Q^{2} \,
\frac{\check{J}^{h(0)}_{-}(x)}{J^{(0)}_{h}}
\label{eq:66ze}
\end{equation}
for the total (equilibrium plus off-equilibrium) current spin polarization.

Turning now to the magnetoresistance of a homogeneous semiconductor, we find
from Eq.~(\ref{eq:80lala}) that the reduced resistance $\tilde{\chi}_{h}$
equals that for  $E_{c}(x) \equiv \Delta(x) \equiv 0$,
\begin{equation}
\tilde{\chi}_{h} = \frac{2l + S}{2l}
\label{eq:66lala}
\end{equation}
[see Eq.~(2.69) of Ref.~\onlinecite{lip05}]. Therefore, we have, from
Eq.~(\ref{eq:80cia}),
\begin{equation}
\tilde{R}_{h} =  \frac{Q}{{\cal G}_{h}^{(0)}} \,  \frac{2l+S}{2l}
\label{eq:66zig}
\end{equation}
for the equilibrium resistance, and, from Eq.~(\ref{eq:66ii}),
\begin{equation}
\check{R}_{h} = - \frac{P }{2 \beta e^{2} J}
 \,  (\alpha_{1} -  \alpha_{2})
\label{eq:66zbg}
\end{equation}
for the off-equilibrium resistance.

\subsection{Current spin polarization in DMS/NMS/DMS heterostructures}

\begin{figure}[t]
\includegraphics[width=0.7\textwidth]{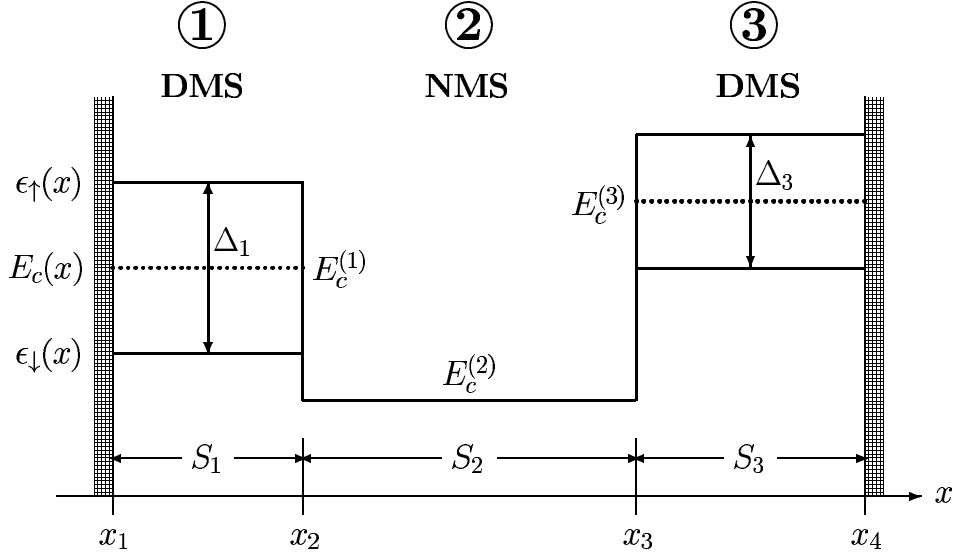}
\caption{Schematic energy diagram for a DMS/NMS/DMS heterostructure composed
of three homogeneous layers.
}
\label{fig:2}
\end{figure}

For parameterizing a DMS/NMS/DMS heterostructure, we attach labels $j=1,2,3$ to
quantities referring to the left DMS layer, the NMS layer, and the right DMS
layer, respectively, and denote the positions of the interfaces (including
the DMS/contact interfaces) by $x_{k}$ $(k=1,2,3,4)$ [see Fig.~\ref{fig:2}],
thereby deviating from the notation used in the main body of the paper.
Quantities referring to the left and right contact are labelled $"l"$ and
$"r"$, respectively.

The position dependence of the polarization inside the left and right contacts
(there labelled 1 and 2) is given by Eq.~(\ref{eq:66ikea}), with
$\mu_{-}(x_{1,4}) = \alpha_{1,4}/\beta$.  The current spin polarization
$P_{J}(x)$ in the three layers is found in terms of the boundary values of the
spin transport function $A(x)$ at the interface positions $x_{k}$, using
Eq.~(\ref{eq:66ze}) together with Eqs.~(\ref{eq:90a}) and (\ref{eq:66zf}),
separately in each layer.  In each of the layers $j$, the function $A(x)$ has
the Sharvin discontinuities $\Delta A_{j}$ and $\Delta A_{j+1}$ at the
interface positions $x_{j}$ and $x_{j+1}$, respectively [see Eqs.~(3.61) and
(3.62) of Ref.~\onlinecite{lip05}].  However, since in the zero-bias limit, $J
\rightarrow 0$, these discontinuities vanish with $J$, we may regard the spin
transport function $A(x)$ as continuous all across the heterostructure.
Recalling that in the zero-bias limit the spin-relaxed chemical
potential $\tilde{\mu}(x)$ may be taken as constant, $\tilde{\mu}(x) \equiv
\tilde{\mu}_{0}$, it is seen from Eq.~(\ref{eq:62}) that the function
$\alpha_{-} (x)$ is also continuous across the entire heterostructure.

In terms of the quantities $\alpha_{k} = \alpha_{-}(x_{k})$ and $A_{k} =
A(x_{k}) = e^{\beta \tilde{\mu}_{0}} \alpha_{k}$, the complete position
dependence of the current spin polarization can then be written as
\begin{equation}
P_{J}(x) = \mp \hat{G}_{l,r} \frac{v_{e} n_{1,4}^{(0)}}{2 J} \alpha_{1,4} \,
e^{-|x_{1,4} - x|/L_{s}^{(l,r)}} \, ,
\label{eq:a1}
\end{equation}
if $x < x_{1}$ and  $x > x_{4}$, respectively, and as
\begin{equation}
P_{J}(x) = P_{j} + \frac{v_{e} N_{c} Q_{j}^{2}}{J_{j}^{(0)}} \gamma_{j}
[C_{j1} e^{-(x - x_{j})/L_{j}} - C_{j2} e^{-(x_{j+1} -
x)/L_{j}}] \, ,
\label{eq:a2}
\end{equation}
if $x_{j} \leq x \leq x_{j+1}$.  Here,
\begin{equation}
\hat{G}_{l,r} = \frac{G_{l,r}}{{\cal G}_{l,r}^{(0)}} \, ,
\label{eq:a3}
\end{equation}
where $G_{l,r}$ is given by Eq.~(\ref{eq:66bbb}), and ${\cal G}_{l,r}^{(0)}$ is
the Sharvin interface conductance (\ref{eq:66zog}).  For the coefficients
$C_{j1}$ and $C_{j2}$, we have \cite{lip05}
\begin{equation}
C_{j1} = \frac{1}{D_{j}} [(1 + \gamma_{j}) e^{S_{j}/L_{j}} A_{j} - (1 -
\gamma_{j}) A_{j+1}]
\label{eq:a4}
\end{equation}
and
\begin{equation}
C_{j2} = - \frac{1}{D_{j}} [(1 - \gamma_{j}) A_{j} - (1 + \gamma_{j})
e^{S_{j}/L_{j}} A_{j+1}] \, ,
\label{eq:a5}
\end{equation}
respectively, where
\begin{equation}
D_{j} = 2 [(1 + \gamma_{j}^{2}) \sinh(S_{j}/L_{j}) + 2 \gamma_{j}
\cosh(S_{j}/L_{j})]
\label{eq:a7}
\end{equation}
and
\begin{equation}
\gamma_{j} = \frac{\bar{l}_{j}}{L_{j}} = \frac{L_{j}}{l_{s}^{(j)}} \, ,
\label{eq:a6}
\end{equation}
with $L_{j}$ and $\bar{l}_{j}$ defined by Eqs.~(\ref{eq:90b}) and
(\ref{eq:90bb}), respectively, and $S_{j} = x_{j+1} - x_{j}$ is the thickness
of layer $j$.

Conservation of the total current $J$ across the heterostructure takes the form
\begin{equation}
J_{j} = \frac{e^{- \beta E_{c}^{(j)}}}{Q_{j}} \, J_{j}^{(0)} = J \; ;
\; j =1,2,3 \, .
\label{eq:a6lala}
\end{equation}
We now require the current spin polarization $P_{J}(x)$ to be continuous at all
interfaces, setting $P_{2} = 0$ (so that $Q_{2} = 1$). This leads to the
following set of four coupled linear equations for the quantities $\alpha_{k}$,
\begin{equation}
(\hat{G}_{l} + Q_{1} g_{1}) \alpha_{1} - Q_{1} h_{1} \alpha_{2} = -
\frac{2J}{v_{e} n_{1}^{(0)}} \, P_{1} \, ,
\label{eq:a8}
\end{equation}
\begin{equation}
- Q_{1} h_{1} \alpha_{1} + \left( Q_{1} g_{1} + e^{\beta
\delta_{12}} g_{2} \right) \alpha_{2} - e^{\beta \delta_{12}} h_{2} \alpha_{3}
= \frac{2 J}{v_{e} n_{1}^{(0)}} \, P_{1} \, ,
\label{eq:a9}
\end{equation}
\begin{equation}
- e^{\beta \delta_{32}} h_{2} \alpha_{2} + ( e^{\beta
\delta_{32}} g_{2} + Q_{3}  g_{3} ) \alpha_{3} - Q_{3}  h_{3}
\alpha_{4}  = - \frac{2 J}{v_{e} n_{4}^{(0)}} \, P_{3} \, ,
\label{eq:a10}
\end{equation}
\begin{equation}
- Q_{3} h_{3} \alpha_{3} + ( \hat{G_{r}} + Q_{3} g_{3})  \alpha_{4} =
\frac{2 J}{v_{e} n_{4}^{(0)}} \,P_{3} \, .
\label{eq:a11}
\end{equation}
Here,
\begin{equation}
g_{j} = h_{j} [\cosh(S_{j}/L_{j}) + \gamma_{j} \sinh(S_{j}/L_{j})]
\label{eq:a12}
\end{equation}
and
\begin{equation}
h_{j} = \frac{4 \gamma_{j}}{D_{j}} \, ,
\label{eq:a13}
\end{equation}
and we have introduced the band offsets
\begin{equation}
\delta_{jj'} = E_{c}^{(j)} - E_{c}^{(j')} \, .
\label{eq:a13a}
\end{equation}
The system of equations (\ref{eq:a8})--(\ref{eq:a11}) is easily solved, so
that the complete position dependence of the current spin
polarization is obtained in explicit form.

Specializing to the case of a {\em symmetric} heterostructure, where the
parameters of the right DMS layer and contact are identical to those
of the left DMS layer and contact, and assuming infinitely high
conductivity of the metal contacts, $\sigma_{l,r} \rightarrow \infty$, so
that $G_{l,r} \rightarrow \infty$, we find
\begin{equation}
\alpha_{1} = \alpha_{4} = 0 \, ,
\label{eq:a14}
\end{equation}
\begin{equation}
\alpha_{2} = - \alpha_{3} = \frac{2 J}{v_{e} n_{D}^{(0)}}
\frac{P}{Q g_{D} + (g_{N} + h_{N}) e^{\beta \delta_{DN}}} \, ,
\label{eq:a15}
\end{equation}
and
\begin{equation}
\hat{G}_{l} \alpha_{1} = - \hat{G}_{r} \alpha_{4} = - \frac{2 J}{v_{e}
n_{D}^{(0)}} \, P + Q h_{D} \alpha_{2} \, .
\label{eq:a16}
\end{equation}
Here, the DMS parameters have been labelled by "D", the NMS parameters by "N",
and $P = P_{1} = P_{3}$.

Thus, in the symmetric case, the current spin polarization $P_{J}(x)$ is
completely determined by the quantity $\alpha_{2}$. Explicitly, we have,
setting $L_{s}^{(c)} = L_{s}^{(l)} = L_{s}^{(r)}$,
\begin{equation}
P_{J}(x) = P \left[ 1 - \frac{Q h_{D}}{Q g_{D} + (g_{N} + h_{N}) e^{\beta
\delta_{DN}} } \right] e^{-|x - x_{1,4}|/L_{s}^{(c)}} \, ,
\label{eq:aa1}
\end{equation}
if $x < x_{1}$ and  $x > x_{4}$, respectively, while
\begin{equation}
P_{J}(x) = P \left\{ 1 - \frac{Q h_{D}}{Q g_{D} + (g_{N} +
h_{N}) e^{\beta \delta_{DN}}} \,[ \cosh (|x - x_{1,4}|/L_{D} ) + \gamma_{D}
\sinh ( |x - x_{1,4}|/L_{D} )] \right\} ,
\label{eq:aa2}
\end{equation}
if $x_{1} \leq x \leq x_{2}$ and $x_{3} \leq x \leq x_{4}$, respectively, and
\begin{equation}
P_{J}(x) = \frac{2P h_{N}}{Q g_{D} e^{- \beta \delta_{DN}} + g_{N} + h_{N} } \,
[ \cosh (S_{N}/2 L_{N}) + \gamma_{N} \sinh (S_{N}/2 L_{N}) ]
\, \cosh ( [x - (x_{2} + x_{3})/2]/L_{N} ) \, ,
\label{eq:aa3}
\end{equation}
if $x_{2} \leq x \leq x_{3}$. Note that, as a consequence of our assumption of
infinitely high conductivity of the contacts, the polarization $P_{J}(x)$ is
independent of the effective electron mass $m^{\ast}$ and of the densities
$n_{D,N}^{(0)}$.

\subsection{Magnetoresistance of DMS/NMS/DMS heterostructures}

The total magnetoresistance of a DMS/NMS/DMS heterostructure, $R = \tilde{R} +
\check{R}$,  is obtained by adding the equilibrium, $\tilde{R}$, and
off-equilibrium, $\check{R}$, contributions corresponding to the three layers
[see Eqs.~(\ref{eq:66zig}) and (\ref{eq:66zbg})],
\begin{equation}
\tilde{R} = \sum_{j=1}^{3}  \frac{Q_{j}}{{\cal G}_{j}^{(0)}} \, \frac{2l_{j}
+ S_{j}}{2 l_{j}}
\label{eq:a17}
\end{equation}
and
\begin{equation}
\check{R} = - \frac{1}{2 \beta e^{2} J} [ P_{1} (\alpha_{1} - \alpha_{2}) +
P_{3} (\alpha_{3} - \alpha_{4}) ] \, .
\label{eq:aa17}
\end{equation}
For a symmetric structure and infinite conductivity in the contacts, we have
\begin{equation}
\tilde{R} =  \frac{2 Q}{{\cal G}_{D}^{(0)}} \, \frac{2l_{D} + S_{D}}{2  l_{D}}
+ \frac{1}{{\cal G}_{N}^{(0)}}\frac{2l_{N} + S_{N}}{2 l_{N}}
\label{eq:a18}
\end{equation}
and, using Eqs.~(\ref{eq:a14}) and (\ref{eq:a15}),
\begin{equation}
\check{R} = \frac{P \alpha_{2}}{\beta e^{2} J} = \frac{1}{{\cal G}_{D}^{(0)}}
\, \frac{2 P^{2}}{ Q g_{D} + (g_{N} + h_{N}) e^{\beta \delta_{DN}}} \, .
\label{eq:a18a}
\end{equation}
We then have
\begin{equation}
\frac{\check{R}}{\tilde{R}} = \frac{2 P^{2}}{Q g_{D} + (g_{N} + h_{N})
e^{\beta \delta_{DN}}} \; \frac{1}{2 Q (2l_{D} + S_{D})/2
l_{D} + e^{- \beta \delta_{DN}}(2l_{N} + S_{N})/2 l_{N}}
\label{eq:a20}
\end{equation}
for the relative magnetoresistance $\check{R}/\tilde{R} \equiv (R -
\tilde{R})/\tilde{R}$. As we have assumed $l_{D,N}$ and $l_{s}^{(D,N)}$ to be
independent of the external magnetic field, expression (\ref{eq:a20}) reflects
solely the "spin accumulation part" \cite{kha05} of the magnetoresistance.
While $\tilde{R}$ and $\check{R}$ depend on the effective electron mass $m^{*}$
and the densities $n_{D,N}^{(0)}$ via the Sharvin interface conductances ${\cal
G}_{D,N}^{(0)}$, the relative magnetoresistance $\check{R}/\tilde{R}$ is
independent of $m^{*}$ and $n_{D,N}^{(0)}$.

\subsection{Diffusive limit}

It is instructive to consider in some detail the diffusive limit of the
thermoballistic description of spin-polarized transport in DMS/NMS/DMS
heterostructures as it allows comparison to the results of
Ref.~\onlinecite{kha05} obtained within the standard drift-diffusion approach
to electron transport.

In the diffusive limit, where $l_{j} \ll S_{j}$ and $l_{j} \ll l_{s}^{(j)}$, we
have $L_{j} \rightarrow L_{s}^{(j)} = \sqrt{l_{j} l_{s}^{(j)}}$ and $\gamma_{j}
\rightarrow l_{j}/L_{s}^{(j)} = \sqrt{l_{j} / l_{s}^{(j)}} \ll 1$ [see
Eqs.~(\ref{eq:90b})--(\ref{eq:90bbb}) and (\ref{eq:a6})],  and hence from
Eqs.~(\ref{eq:a12}) and (\ref{eq:a13})
\begin{equation}
g_{j} = h_{j} \, \cosh(S_{j}/L_{s}^{(j)})
\label{eq:a21}
\end{equation}
and
\begin{equation}
h_{j} = \frac{2 l_{j}}{L_{s}^{(j)}} \, \frac{1}{\sinh (S_{j}/L_{s}^{(j)})} \, ,
\label{eq:a22}
\end{equation}
respectively. Equation (\ref{eq:a15}) for the symmetric heterostructure then
reduces to
\begin{equation}
\alpha_{2} = \frac{2 \beta e^{2} J P}{Q^{2} G_{D}
\coth(S_{D}/L_{s}^{(D)}) + G_{N} \coth(S_{N}/2L_{s}^{(N)}) \, e^{\beta
\delta_{DN}}} \, , \label{eq:a23}
\end{equation}
where
\begin{equation}
G_{j} = \frac{\sigma_{j}}{L_{s}^{(j)}}
\label{eq:a23a}
\end{equation}
($j = D,N$), and  $\sigma_{j}$ is the (spin-summed) conductivity,
\begin{equation}
\sigma_{j} = 2 \beta e^{2} v_{e} \tilde{n}_{j} l_{j} = \frac{2}{Q_{j}} {\cal
G}_{j}^{(0)} l_{j} .
\label{eq:a16a}
\end{equation}
Now, using Eq.~(\ref{eq:a23}) with $\delta_{DN} = 0$ in Eq.~(\ref{eq:a2}), we
find that the values of the current spin polarization $P_{J}(x)$ at $x_{1}$,
$x_{2}$, and $(x_{2} + x_{3})/2$, respectively, agree with those given by
Eqs.~(16)--(18) of Ref.~\onlinecite{kha05}, if we identify in the latter
equations the spin-flip lengths $\lambda_{D}$ and $\lambda_{N}$ with
$L_{s}^{(D)}$ and $L_{s}^{(N)}$, respectively, and the layer thicknesses $d$
and $2 x_{0}$ with $S_{D}$ and $S_{N}$, respectively.

Similarly, from Eq.~(\ref{eq:a18}) and from Eq.~(\ref{eq:a18a}), using
Eq.~(\ref{eq:a23}), we find the equilibrium contribution to the
magnetoresistance, $\tilde{R}$, and the off-equilibrium contribution,
$\check{R}$, respectively, to agree with the corresponding results of
Ref.~\onlinecite{kha05}, {\em viz.,} the first and second term on the
right-hand side of Eq.~(10) of that reference.  Then, trivially, we find
agreement of the relative magnetoresistance $\check{R}/\tilde{R}$ given by
Eq.~(\ref{eq:a20}) with the first term on the right-hand side of Eq.~(13) of
Ref.~\onlinecite{kha05}.

\subsection{Ballistic limit}

In the ballistic limit, where $l_{j} \gg S_{j}$ and $l_{j} \gg l_{s}^{(j)}$, we
have $L_{j} \rightarrow l_{s}^{(j)}$ and $\gamma_{j} \rightarrow 1$ [see
Eqs.~(\ref{eq:90b})--(\ref{eq:90bbb}) and (\ref{eq:a6})], so that
$g_{j} = 1$ and $h_{j} = e^{- S_{j}/l_{s}^{(j)}}$.

For the symmetric heterostructure, we then have from Eq.~(\ref{eq:a15})
\begin{equation}
\alpha_{2} = \frac{2 J}{n_{N}^{(0)}} \frac{P}{Q e^{- \beta \delta_{DN}} + 1 +
e^{-S_{N}/l_{s}^{(N)}}}
\, .
\label{eq:a21a}
\end{equation}
Using this expression for $\alpha_{2}$, we obtain the ballistic limit for the
current spin polarization and the magnetoresistance.  The equilibrium
contribution to the magnetoresistance, in particular, follows from
Eq.~(\ref{eq:a18}) in the form
\begin{equation}
\tilde{R} = 2 \, \frac{Q}{{\cal G}_{D}^{(0)}} + \frac{1}{{\cal G}_{N}^{(0)}}
\, ,
\label{eq:22}
\end{equation}
in which the Sharvin interface resistances $1/{\cal G}_{D,N}^{(0)}$
characterize, via their dependence on the constant densities $n_{D,N}^{(0)}$,
the (homogeneous) DMS and NMS layers.  Notably, owing to the presence of the
factor $Q$, the contribution to $\tilde{R}$ from the DMS layers vanishes when
$P \rightarrow 1$.  If we had not assumed the interfaces at $x = x_{2,3}$ to be
fixed points of thermal equilibrium, the ballistic equilibrium
magnetoresistance would be given in terms of the Sharvin interface resistances
at the end-points $x_{1,4}$ [see Eq.~(2.51) of Ref.~\onlinecite{lip05}] and of
the potential barriers provided by the profiles (\ref{eq:01}).

\section{DMS/NMS/DMS heterostructures:\ Application}

\begin{figure}[t]
\includegraphics[width=0.7\textwidth]{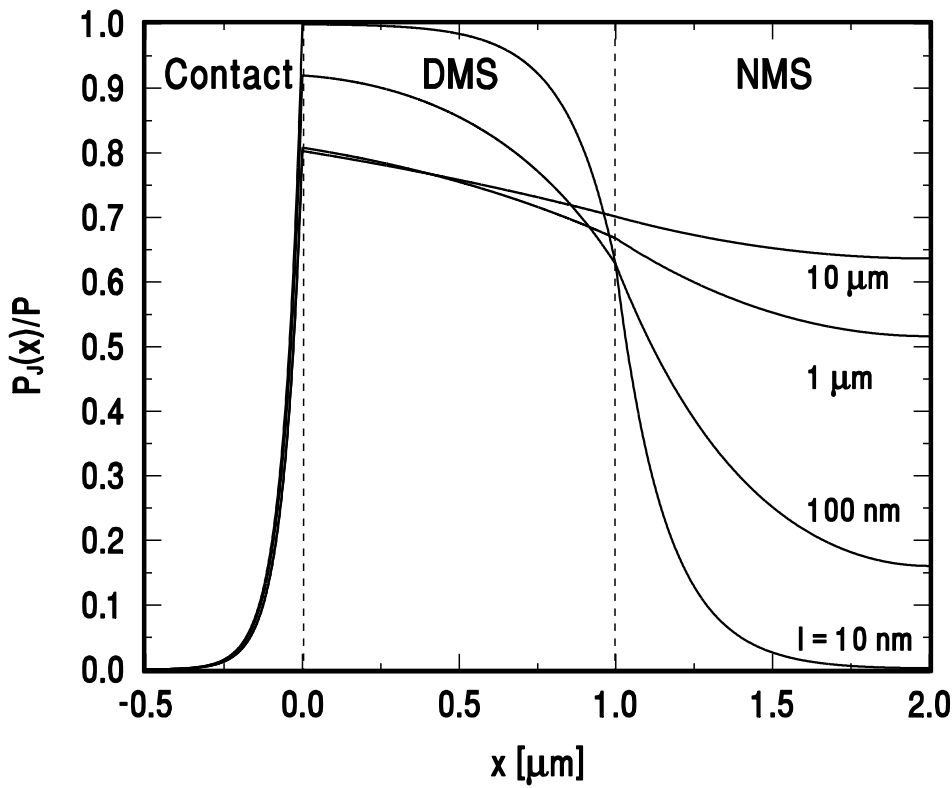}
\caption{
Relative zero-bias current spin polarization $P_{J}(x)/P$ across a symmetric
DMS/NMS/DMS heterostructure enclosed between metal contacts with infinitely
high conductivity, calculated from Eqs.~(\ref{eq:aa1})--(\ref{eq:aa3}) for zero
band offset, $\delta = 0$.  The different curves correspond to different values
of the momentum relaxation length $l = l_{D} = l_{N}$.  The remaining parameter
values are:\ $S_{D} = 1$ $\mu$m, $S_{N} = 2$ $\mu$m, $l_{s}^{(D)} = l_{s}^{(N)}
= 2.5$ $\mu$m, $P = 0.8$, $L_{s}^{(c)} = 60$ nm.
}
\label{fig:3}
\end{figure}

In the thermoballistic description of spin-polarized electron transport in
DMS/NMS/DMS heterostructures, the momentum and ballistic spin relaxation
lengths, $l$ and $l_{s}$, respectively, are the fundamental dynamical
quantities. These, apart from the potential profiles $E_{c}^{(D,N)}$ and the
geometric dimensions $S_{D,N}$, determine the current spin polarization and the
magnetoresistance of the structure as a function of the Zeeman splitting
$\Delta$, i.e., of the strength of the external magnetic field.  Here, we
illustrate the dependence of polarization and magnetoresistance on $l$ by
specific numerical examples.  (We recall \cite{lip05} that, when varying $l$,
we understand the different values to represent a {\em class} of semiconductors
which have similar material properties, but differ in the strength of impurity
and phonon scattering.)  Furthermore, we point out the possibility of
determining (experimentally) the quantities $l$ and $l_{s}$ from their effect
on the (experimentally accessible) spin polarization and magnetoresistance.

In Fig.~\ref{fig:3}, the relative current spin polarization, i.e., the ratio of
current spin polarization and static DMS polarization, $P_{J}(x)/P$, across a
symmetric DMS/NMS/DMS heterostructure at zero bias is shown for different
values of the momentum relaxation length $l = l_{D} = l_{N}$ at a fixed value
of the ballistic spin relaxation length $l_{s} = l_{s}^{(D)} = l_{s}^{(N)}$.
The behavior of $P_{J}(x)/P$ exhibits two features.

First, the injected spin polarization, i.e., the value of the polarization at
the DMS/NMS interfaces, remains on a high, weakly $l$-dependent level ranging
between 0.6 and 0.7 when $l$ varies by three orders of magnitude from the
diffusive to the ballistic regime.  This behavior contrasts with that
calculated within the thermoballistic treatment of spin injection at the
interfaces of the  metallic ferromagnets and the semiconductor in a
ferromagnet/NMS/ferromagnet structure \cite{lip05}. The results for the latter
case are shown in Fig.\ 5 of Ref.~\onlinecite{lip05}.  These have been obtained
with an $l_{s}$-value equal to that used in Fig.~\ref{fig:3} of the present
paper, and with the bulk polarization of the ferromagnets equal to the static
DMS polarization.  The injected polarization at the ferromagnet/semiconductor
interfaces (for zero interface resistance) is seen to be very small in the
diffusive limit, and rapidly rising with increasing $l$.  The strong dependence
on $l$ reflects the conductivity mismatch \cite{sch00} between metallic
ferromagnets and semiconductor, which is very large in the diffusive (low-$l$)
regime where the conductivity of the semiconductor vanishes with $l$.  No such
mismatch occurs at the DMS/NMS interfaces in a DMS/NMS/DMS heterostructure.

Second, the behavior of the polarization inside the NMS layer is governed by
the $l$-dependence of the generalized spin diffusion length $L = [l l_{s}/(1 +
l/l_{s})]^{1/2}$.  In the diffusive limit, where $L \approx \sqrt{l l_{s}}$,
the polarization dies out rapidly; in the ballistic limit, we have $L \approx
l_{s}$, i.e., the decay is determined solely by the ballistic spin relaxation
length.  Qualitatively, the behavior of the polarization inside the NMS layer
does not depend on the kind of spin injector (ferromagnet or DMS).  However,
due to the larger injected polarization, the magnitude of the polarization
inside the NMS layer is larger for the DMS injector.

\begin{figure}[t]
\includegraphics[width=0.7\textwidth]{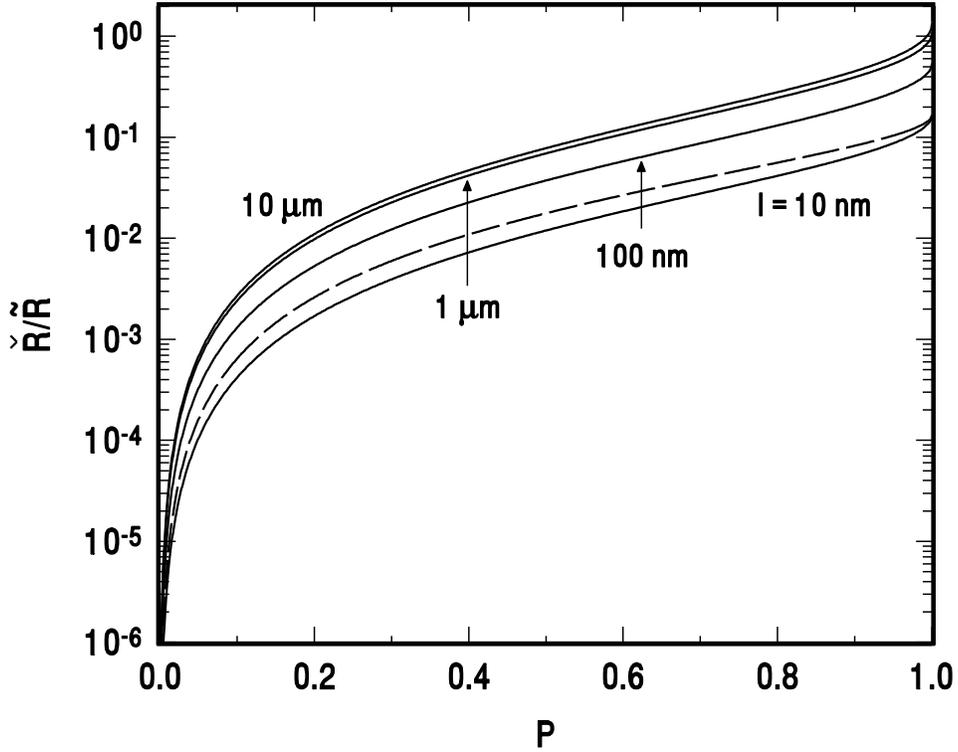}
\caption{Relative magnetoresistance $\check{R}/\tilde{R}$ of
a symmetric DMS/NMS/DMS heterostructure as a function of
the static DMS  polarization $P$, calculated from Eq.~(\ref{eq:a20}) with the
parameter values of Fig.~\protect{\ref{fig:3}} (solid curves). The long-dashed
curve for $l = 10$ nm has been calculated with $S_{D} = 0.2$ $\mu$m.
}
\label{fig:4}
\end{figure}

In Fig.~\ref{fig:4} (solid curves), the relative magnetoresistance $\check{R}/
\tilde{R}$ of a symmetric DMS/NMS/DMS heterostructure is shown as a function of
the static DMS polarization $P$ for the parameter values of Fig.~\ref{fig:3}.
For $P$ not too close to unity, the qualitative behavior of the curves is
determined by the overall factor $P^{2}$ in Eq.~(\ref{eq:a20}).  When $P$
approaches unity, the curves rise sharply owing to the $P$-dependence of the
terms proportional to $Q$ in the denominator of the expression for $\check{R}/
\tilde{R}$.  As a function of $l$, $\check{R}/\tilde{R}$ rises, over the full
$P$-range, by about one order of magnitude when $l$ varies from the diffusive
to the ballistic regime.

In the example of Fig.~\ref{fig:4}, $\check{R}/ \tilde{R}$ tends to increase
with {\em decreasing} thickness $S_{D}$ of the DMS layers, as is seen by
comparing the solid and long-dashed curves for $l = 10$ nm (the magnitude of
this increase becomes progressively smaller with increasing $l$).  In the
diffusive calculation of Ref.~\onlinecite{kha05}, the relative
magnetoresistance was found, for a specific parameter choice, to increase with
{\em increasing} $S_{D}$.  We have reproduced this behavior quantitatively by
choosing the parameter values in Eq.~(\ref{eq:a20}) so as to conform to those
underlying the results shown in Fig.~3b of Ref.~\onlinecite{kha05}.  In
particular, we took $l_{D} = 10$ nm and used Eqs.~(\ref{eq:90bbb}) and
(\ref{eq:a16a}), respectively, to relate the ballistic spin relaxation lengths
$l_{s}^{(D,N)}$ to the spin diffusion lengths $\lambda_{D,N}$ of
Ref.~\onlinecite{kha05}, and the ratio $l_{D}/l_{N}$ to the ratio
$\sigma_{D}/\sigma_{N}$.  A detailed study of the full parameter dependence of
the magnetoresistance is clearly desirable.

For spin-polarized electron transport in DMS/NMS/DMS heterostructures, the
thermoballistic description in the simplified form presented in this paper
leads to transparent explicit expressions for the current spin polarization
[Eqs.~(\ref{eq:aa1})--(\ref{eq:aa3})] and for the magnetoresistance
[Eqs.~(\ref{eq:a18}) and (\ref{eq:a18a})].  If these turn out to reproduce,
albeit approximately, the results of actual measurements for different values
of $P$ and various choices of $S_{D} $ and $S_{N}$, one should be able to
determine the values of the parameters $l$ and $l_{s}$ for magnetic and
nonmagnetic materials from experiment in a quite direct manner.

\section{Summary and conclusions}

We have presented the systematic extension of the thermoballistic description
of spin-polarized electron transport in semiconductors to the case of a
spin-split conduction band, allowing us to envisage applications to
spin-polarized transport in paramagnetic DMS and in DMS/NMS/DMS
heterostructures.

Assuming arbitrarily shaped potential profiles exhibiting arbitrary,
position-dependent spin splitting, we have constructed the thermoballistic
currents and densities by starting from the spin-resolved densities at points
of local thermal equilibrium, at which electron currents across ballistic
transport intervals are activated.  These currents are subject to spin
relaxation.  Dividing the ballistic currents and densities into their
equilibrium (spin-relaxed) and off-equilibrium parts, we have obtained a
description of the equilibrium parts in terms of the spin-relaxed chemical
potential, and of the off-equilibrium parts in terms of a spin transport
function that is related to the splitting of the spin-resolved chemical
potentials.  From the ballistic currents and densities, the corresponding
thermoballistic quantities are constructed by weighted summation over all
ballistic intervals.  The procedures for calculating the spin-relaxed chemical
potential and the spin transport function are outlined.

The thermoballistic description of spin-polarized electron transport has been
applied to the calculation of the current spin polarization and
magnetoresistance in DMS/NMS/DMS heterostructures. The results have been
compared to those of the customary drift-diffusion approach. The important role
of the fundamental momentum and spin relaxation lengths as well as the
possibility of their experimental determination are pointed out.

The emphasis in the present paper has been on the systematic development of the
formalism.  In future work, this formalism will have to be implemented in
detail.  To this end, a careful study dealing with the role of the interfaces
in electron transport in heterostructures, as well as their adequate modelling
within the thermoballistic description, is required.  Furthermore, efficient
algorithms for the solution of the integral equations for the resistance
function and the spin transport function are to be developed.  Then, by
performing exploratory calculations sampling the full parameter space, one may
be able to identify novel features and trends in spin-polarized electron
transport in paramagnetic semiconductors.

\end{document}